\documentclass[%
reprint,
superscriptaddress,
showpacs,
amsmath,amssymb,
pre,
]{revtex4-1}

\usepackage{graphicx}
\usepackage{dcolumn}
\usepackage{bm}
\usepackage{color}

\usepackage[retainorgcmds]{IEEEtrantools}

\newcommand{\aj}{AJ}

\newcommand{\icarus}{ICARUS}
\newcommand{\celmech}{Celest.~Mech.~Dyn.~Astron.}

\newcommand{\pr}{Phys.~Rev.} 
\newcommand{\physrep}{Phys.~Rep.} 
\newcommand{\physicaD}{Physica D}

\newcommand{\jpa}{J.~Phys.~A - Math.~Gen.}
\newcommand{\ajp}{Am.~J.~Phys.}

\newcommand{\dcdsB}{Disc.~Cont.~Dyn.~Syst.--B}
\newcommand{\russmath}{Russ.~Math.~Surv.}
\newcommand{\theoryprobappl}{Theory Probab.~Appl.}
\newcommand{\cmp}{Commun.~Math.~Phys.}

\begin{document}

\preprint{APS/123-QED}

\title{Stochastic approach to diffusion inside the chaotic layer of a resonance}

\author{Mart{\'i}n F. Mestre}
\email{mmestre@fcaglp.unlp.edu.ar}
\affiliation{%
  Grupo de Caos en Sistemas Hamiltonianos.\\
  Facultad de Ciencias Astron\'omicas y Geof\'isicas, UNLP, Argentina.\\
  Instituto de Astrof\'isica de La Plata (CCT La Plata - CONICET, UNLP), Argentina.\\
}%

\author{Armando Bazzani}
\affiliation{Dipartimento di Fisica, Universit{\`{a}} di Bologna, Italia.\\
  INFN sezione di Bologna, Italia.
}%
\author{Pablo M. Cincotta}
\affiliation{%
  Grupo de Caos en Sistemas Hamiltonianos.\\
  Facultad de Ciencias Astron\'omicas y Geof\'isicas, UNLP, Argentina.\\
  Instituto de Astrof\'isica de La Plata (CCT La Plata - CONICET, UNLP), Argentina.\\
}%

\author{Claudia M. Giordano}
\affiliation{%
  Grupo de Caos en Sistemas Hamiltonianos.\\
  Facultad de Ciencias Astron\'omicas y Geof\'isicas, UNLP, Argentina.\\
  Instituto de Astrof\'isica de La Plata (CCT La Plata - CONICET, UNLP), Argentina.\\
}%

\date{\today}

\begin{abstract}
  We model chaotic diffusion, in a symplectic 4D map by using the result of a theorem that was developed
  for stochastically perturbed integrable Hamiltonian systems.
  We explicitly consider a map defined by a free rotator (FR) coupled to a standard map (SM).
  We focus in the diffusion process in the action, $I$, of the FR, obtaining a semi--numerical method to compute the
  diffusion coefficient. We study two cases corresponding to a thick and a thin chaotic layer in the 
  SM phase space and we discuss a related conjecture stated in the past.
  In the first case the numerically computed probability density function for the action
  $I$ is well interpolated by the solution of a Fokker-Planck (F-P) equation, whereas it presents
  a non--constant time delay respect to the concomitant F-P solution in the second case suggesting the presence of
  an anomalous diffusion time scale.
  The explicit calculation of a diffusion coefficient for a 4D symplectic
  map can be useful to understand the slow diffusion observed in Celestial Mechanics and Accelerator Physics.
\end{abstract}

\maketitle

{\color{blue}
{\small{This is the author's version of a work that was submitted to Physical Review E (http://pre.aps.org).}}
}

\section{\label{sec:introduction}Introduction}

\noindent Diffusion in Hamiltonian systems with more than 2DoF is a
long--standing open problem whose understanding is relevant to model
the slow diffusion phenomena observed in physical systems well
described by conservative deterministic differential equations.
Ref.~\cite{2001MORBIDELLI_RegChaoticDyn} explains the concept of
chaotic diffusion from the point of view of Celestial Mechanics and
gives examples of this relevant process for the dynamics of small
bodies of the solar system. Ref.~\cite{2007LNP...729..111T} gives
further evidence of the fact that chaotic diffusion is an important
element of the long--term dynamics of the asteroid belt and shows
how simple models of chaotic diffusion can be used to estimate the
age of asteroid families. Ref.~\cite{2008Icar..196....1L} makes a
statistical study of the stability of the solar system by computing
probability density functions (PDFs) for the eccentricity and
inclination of the planets over 5 Gyr. Further applications of
chaotic diffusion in the Celestial Mechanics setting are described
in Refs.~\cite{2005dpps.conf..157V,1998AJ....116.3029N}. Another
relevant application is the transverse diffusion in circular
accelerators when one takes into account the multipolar components of the magnetic field
 which limit the dynamics aperture\cite{1998Bazzani_proceeding}. The coupling of the transverse phase space with
longitudinal particle dynamics, the presence of scattering and
parametric dependence due to supplied current modulations (ripples)
have been proposed as possible causes of the particle
diffusion\cite{mais1999}. A possible explanation of the underlying
diffusion mechanism is the simultaneous presence of a stochastic or
chaotic perturbation and the nonlinear
terms\cite{1992CMaPh.149...97C}. Ref.~\cite{1994PhyD...71..122B}
carries out a numerical computation of the slow diffusion of orbits through thin chaotic
layers of a 4D symplectic map that models accelerator dynamics.
Recently, experiments have been
performed to measure the beam diffusion rate in proton
colliders\cite{stancari2011} and an explicit calculation of the
diffusion coefficients for 4D
stochastically perturbed maps is needed to analyze the experimental data.\par
In the case of nearly--integrable Hamiltonian systems, the interest is focused in the action variables.
Depending on the strength of the perturbation there are restrictions to the topology of the region of action
space in which global diffusion can take place.
In Refs.~\cite{1977RuMaS..32....1N,1979PhR....52..263C,2005nlin......7059G,springerlink:10.1007/s10569-008-9151-8}
the concept of {\it Arnold diffusion} (AD) is understood as the diffusion that takes place along the
{\it Arnold web}\footnote{{Arnold web} is the intersection of all the perturbed resonant surfaces with the isoenergetic
manifold.} of a system that satisfies simultaneously the hypothesis of
the KAM~\cite{1989ARNOLD_VLADIMIR_BOOK} and the Nekhoroshev~\cite{1977RuMaS..32....1N} theorems.
In this sense, Refs.~\cite{2003PhyD..182..179L,2005nlin......7059G,2005CeMDA..92..243F,2011WSAAA...3..319C,2013PhyD..251...19E,2013arXiv1310.3158C} show numerical evidence of AD.
For particular type of systems, there are theorems focused in
demonstrating analytically the existence of orbits which drift in action space a quantity of order one in a finite
time interval. In particular it has been proved~\cite{2001PhyD..156..201E} that a twist map, coupled to a map close to the anti--integrable limit,
has many orbits that drift arbitrarily far. Their generic system has
a phase space with a geometry similar to the one of the {\it thick layer problem} (see below).
Chirikov's theory~\cite{1979PhR....52..263C} provides a formula to estimate quantitatively
the value of the diffusion coefficient. It has been
tested~\cite{1980AIPC...57..323C} in a particular 2.5DoF nearly integrable Hamiltonian system, obtaining
a good agreement between theory and experiments in a particular range of the size of the perturbation.
The theory of Nekhoroshev~\cite{1977RuMaS..32....1N,2006MORBIDELLI_BOOK} gives upper (but not lower) bounds to the rate of AD.
The implications of this theory to the geometry and speed of AD, in singly and multiply resonant domains in nearly
integrable Hamiltonian systems, have recently been comprehensively reviewed
by Refs.~\cite{springerlink:10.1007/s10569-008-9151-8,2013PhyD..251...19E}. They show more light onto the connection between the
diffusion coefficient and the size of the remainder of the optimal resonant normal form and propose an associated
set of variables in which the AD can be visualized and measured, thus stablishing a novel path for future
stability and diffusion studies (e.g.~\cite{2013CeMDA.117..101E,2013arXiv1310.3158C}).\par
Another theoretical approach considers the {\it stochastic pump}~\cite{1980AIPC...57..272T,NYAS:NYAS119} model.
Both references consider a 4D symplectic map whose phase space contains
a {\it thick} and a {\it thin} chaotic layer, which correspond to
{\it overlapping} and {\it non-overlapping} regimes, respectively.
In their model, the dynamics on the chaotic layer is responsible
for the diffusion of a perturbed action associated to a libration regime.
According to the width of the layer, they apply two analytical procedures to find
that the diffusion coefficient is strongly dependent on the ratio
of the characteristic frequencies of the chaotic layer and the
librations. Nevertheless, due to the unavoidable approximations,
this theory does not allow a detailed analysis of the dependence
of the diffusion behaviour on the local action.
They perform numerical experiments to corroborate their results.
Refs.~\cite{1990PhLA..151...37W,1990PhRvA..42.5885W}
verify this method in a 4D symplectic map
that consists of two coupled standard maps (SM).
Ref.~\cite{1990PhRvA..42.5885W} also estimates the global
rate of diffusion by weighting
local diffusion rates with the relative volume occupied by
the various chaotically accessible regions in the 4D phase space.
This map has also been studied by Refs.~\cite{1985PhLA..110..435K,HONJO_KANEKO},
where the diffusion rate has been measured for different values of the coupling parameter.
The definition of the diffusion coefficient used in~\cite{HONJO_KANEKO} is not based
in an ensemble average but in a time average of a single trajectory.
For other dynamical and theoretical discussions see Refs.~\cite{1994CHIERCHIA_GALLAVOTTI,2009CMaPh.290..557G,1995LOCHAK}.

The development of the Stochastic Dynamical Systems
Theory\cite{freidlin1984} allowed to describe the diffusion in
Hamiltonian systems by means of stochastic perturbations which mimic
the chaotic dynamics \cite{bazzani1997,just2001,baba2006} and to
derive a Fokker-Planck for the probability distribution function
(PDF) in the slow variables\cite{baba2007}. In some cases it was
possible to prove diffusive limit theorems for dynamical systems in
presence of (deterministic) chaotic perturbations letting that the
amplitude of the perturbation tends to zero and the time to
infinity\cite {melbourne2011}. According to this point of view it is
possible to justify averaging principles to prove the diffusive
limit\cite{gottwald2013}. 
In Ref.~\cite{1992_Lichtenberg-Lieberman}, a Fokker-Planck (F-P)
equation for the PDF associated to the action variable of a 1.5DoF
Hamiltonian system is deduced, by means of the random phase
approximation. This approximation is partially valid in regions with
resonance overlap~\cite{1979PhR....52..263C} and is based on the
assumption that the mixing of the angle variables of a Hamiltonian
system is sufficiently faster than the one of the action variables
({\it quasi}--linear theory). In Ref.~\cite{2005VARVOGLIS.book}
there has been applied the {\it quasi}--linear theory to another
1.5DoF Hamiltonian system, obtaining a F-P equation whose diffusion
coefficient turns out to depend only on the action variable.
Ref.~\cite{2010PhRvL.104w5001K} presents a hierarchy of equations
for the evolution of the PDF in the phase space of
nearly--integrable Hamiltonian systems of arbitrary dimension. In
this method, the kinetic equation has time dependent coefficients,
even in the case of an autonomous perturbation. This represents a
major difference with respect to {\it quasi}--linear theories. These
equations have been tested numerically only in a 1.5DoF Hamiltonian
system. Ref.~\cite{2002PhR...371..461Z} reviews many fractional
kinetic models and their relationship with dynamical models, phase
space topology and other chaos characteristics, as Poincar{\'e}
recurrences and sticky domains. Summing up, although there are
previous works that model Hamiltonian diffusion with a F-P equation,
these are mainly focused in systems with less than 2DoF.

Ref.~\cite{1990PhRvA..41.4143K} considers the following {\it a priori}
unstable\footnote{Following the definition given in Sec.~2 of Ref.~\cite{2009CMaPh.290..557G}.}
4D symplectic map:
\begin{IEEEeqnarray}{lcl}
  \label{map_Meiss}
  I_{n+1}&=&I_n  -\epsilon \sin(\theta_n + \psi_n ) \nonumber \\
  \theta_{n+1}&=&\theta_n + I_{n+1} \hspace{1.2cm} mod \hspace{0.1cm} 2\pi \nonumber \\
  J_{n+1}&=&J_n + K\sin(\psi_n)   -\epsilon \sin(\theta_n + \psi_n ) \nonumber \\
  \psi_{n+1}&=&\psi_n + J_{n+1}  \hspace{1.2cm} mod \hspace{0.1cm} 2\pi.
\end{IEEEeqnarray}
For $\epsilon=0$ and $K\neq0$ the map consists of two uncoupled 2D maps: a free rotator
in the $[I,\theta]$--plane and a SM of parameter $K$ in the $[J,\psi]$--plane.
In Ref.~\cite{1990PhRvA..41.4143K} the authors show that in the case in which $\epsilon\neq0$
the short-time correlations in the $[J,\psi]$-plane due to the chaotic layer, affect
the diffusion in the $[I,\theta]$--plane.
They apply the {\it characteristic function method}~\cite{1981PhRvA..23.2744C}
and find that the diffusion tensor depends on the parameters of the system but not on the action ($I$).
They make experiments for wide ranges of parameter values,
finding agreement with predictions as long as $|K|>2$ and $|\epsilon|>2$.
They remark that this system has two interesting limit situations:
$|K|>>1$, $|\epsilon|<<1$ (thick layer) and $|K|\lesssim1$, $|\epsilon|<<1$ (thin layer),
both being out of reach of their method.
Moreover Ref.~\cite{1998JPhA...31.5843B} presents numerical evidence of the fact
that the correlation function of the increments $\Delta J_n$ of a
SM with $K=3$ can be fitted (approximately) by an exponentially
decaying oscillating function, which is the exact autocorrelation
function of the stochastic rotator. The authors conjecture that
\textit{if the chaotic movement of the $\psi$ variable is coupled
with the dynamics of an integrable system, then it could be observed
a diffusion in the phase space of the integrable system which is
similar to the diffusion driven by a stochastic rotator}.

In this paper we will make a numerical application, on the map given by Eq.~(\ref{map_Meiss}),
of the averaging theorem discussed in~\cite{1998JPhA...31.5843B}
to be used in stochastically perturbed nearly--integrable Hamiltonian systems, which generalizes the result\cite{1992CMaPh.149...97C}.
We predict and numerically compute the diffusion coefficient associated to $I$ for two values of the parameter of the SM that correspond
to the cases of thick and thin layer diffusion.
We show that in the thick layer regime the PDF satisfies a F-P equation in a ``slow diffusion time'', while in the
thin layer regime the PDF presents a ``time delay'' with respect to the associated F-P solution in the slow diffusion time,
that could be related to a different scaling law between the real time and the diffusion time when the 
correlation of the chaotic perturbation is not decaying sufficiently fast. 
Moreover, we will review the mentioned conjecture.

In principle, our approach is applicable to any non-integrable Hamiltonian system that can be,
either locally or in the whole phase space, decomposed into an integrable and a chaotic system
which are weakly coupled:
e.g. the case of a perturbed simple nonlinear resonance.
Implementation of the {F-P} equation can facilitate the study of the parametric dependence on the diffusion process for a whole
particle distribution.

The structure of the article is as follows.
In Sec.~\ref{sec:averaging_theorem} we will provide a short description
of a version of the averaging theorem for stochastically perturbed
integrable maps.
In Sec.~\ref{sec:Gaussian_example} we will give an example of its use
with harmonic noise.
In Sec.~\ref{sec:method} we will introduce a semi-numerical method to compute the diffusion coefficient
for the map~(\ref{map_Meiss}).
In Secs.~\ref{sec:thick_chaotic_layer} and ~\ref{sec:thin_chaotic_layer}
we will test the method, respectively, in the thick and thin layer regimes.
Finally, in Sec.~\ref{sec:conclusion} we provide the conclusions.

\section{\label{sec:averaging_theorem} An averaging theorem for stochastic systems}
\noindent
The first averaging theorem for a deterministic equation stochastically perturbed
is established in Ref.~\cite{1966Khasminskii}.
Afterwards, in Ref.~\cite{1992CMaPh.149...97C} this result is generalized
proving that under certain conditions, a first integral of the unperturbed system
{\it weakly}\footnote{Weak convergence stands for convergence of the PDF}
converges towards a diffusion process.
This theorem can be extended
to compute the diffusion limit of the dynamics of the actions of a
stochastically perturbed Hamiltonian system (see \cite{1998JPhA...31.5843B} and references therein)
and it has been applied in the case of an integrable Hamiltonian system perturbed by
a stochastic rotator.
There the authors verify one of the conclusions of the theorem, which
states that for small enough perturbations, the PDF of the action satisfies a F-P equation
whose diffusion coefficient depends on the correlation function of the
stochastic process.

Ref.~\cite{1998Bazzani_proceeding} applies the theorem
to a stochastically perturbed symplectic map. They show that
in the limit of noise with small amplitude, a colored noise
can excite a local diffusion of the action variable.
In what follows we restate the theorem, without demonstration.

Let $M:\mathbb{R}^2 \rightarrow \mathbb{R}^2$ be a symplectic map
with an elliptic fixed point in the origin and let $\mathcal{R}$
be a neighborhood of it that defines a stable region.
Let us assume that in $\mathcal{R}$  the measure of the nonlinear resonances
and of the chaotic regions is negligible so that it is possible to replace the
original map by an integrable one, $M_0:\mathbb{R}^2 \rightarrow \mathbb{R}^2$.
Moreover, let  $\mathbf{x}\equiv (x,p)\in \mathbb{R}^2$.

Let $\xi_n$ be a stationary stochastic process, with zero mean value and unit variance,
defined in some probability space associated to some
sample space, $\mathcal{S}$.
Let $E[\cdot]$ denote the concomitant (theoretical)
expectation value.

We consider a stochastic map, $P_n$, of the following form:
\begin{equation*}
  P_n(\mathbf{x})=\mathbf{x}+\epsilon \xi_n \mathbf{v}(\mathbf{x}),
\end{equation*}
where $n\in \mathbb{N}$ and
\begin{equation*}
  \mathbf{v}(\mathbf{x})=\left(
  \begin{array}{c}
    0\\
    -\frac{d\tilde{V}(x)}{dx}
  \end{array} \right),
\end{equation*}
with $\tilde{V}(x)$ being a potential function.

Then, we will study the dynamics of the stochastically perturbed
symplectic map:
\begin{equation}
  \label{stoch_pert_symp_map}
  \mathbf{x}_{n+1}=P_n\circ M_0 (\mathbf{x}_n),
\end{equation}
where ``$\circ$'' denotes the composition operation.

Introducing the action--angle variables $(I,\theta)$ of the map $M_0$,
the map~(\ref{stoch_pert_symp_map}) is rewritten as:
\begin{equation}
  \label{stoch_pert_symp_map_aa}
  \left(
  \begin{array}{c}
    \theta_{n+1} \\
    I_{n+1}
  \end{array} \right)
  =
  \exp(\epsilon \xi_n L_{V(I,\theta)})
  \circ
   \left(
   \begin{array}{c}
     \theta_{n}+\Omega(I_n) \\
     I_{n}
   \end{array} \right),
\end{equation}
where $V\equiv V(I,\theta)=\tilde{V}(x(I,\theta))$,
$L_V$ is the Lie operator defined by the Poisson bracket
$\{\cdot,V\}$  with the potential function and $\Omega(I)$ is the frequency of $M_0$.

The potential admits a Fourier series development:
\begin{equation}
  \label{Fourier_pot}
  V(I,\theta) =\sum_k V_k(I)\mathrm{e}^{\mathrm{i}k\theta}.
\end{equation}
For a fixed value of $\epsilon$ we introduce a frequency cut--off $k_{max}$ in the expansion~(\ref{Fourier_pot})
in order to neglect those terms with $||V_k||\le \epsilon$, where $||\cdot||$ denotes the supremum, or infinite, norm in $\mathcal{R}$.

The hypothesis of the theorem are the following:
\begin{itemize}
\item The unperturbed map is at least of class $\mathcal{C}^2$ in $\mathcal{R}$.
\item The noise $\xi_n$ satisfies a $\varphi$--mixing condition; i.e. if $f(x)$ and $g(x)$ are
  bounded measurable functions then the following inequality holds:
  \begin{equation*}
    |E[f(\xi_n)g(\xi_0)]-E[f(\xi_n)]E[g(\xi_0)]|\le ||f||~||g||~\varphi(n),
  \end{equation*}
  where the function $\varphi(n)$ is such that $\lim_{n\rightarrow \infty}n^6\varphi(n)=0$.
\item No resonance condition, of the form $k\Omega(I)-2\pi q =0$ ($q\in\mathbb{Z}$)
for $|k|<2 k_{max}$ ($k\neq 0$), is fulfilled in  $\mathcal{R}$ when $V_k(I)\neq 0$
in the expansion~(\ref{Fourier_pot}).
\item The following limit exists:
  \begin{IEEEeqnarray}{ll}
    \label{lim_coef_dif}
    \mathcal{D}(I)\equiv &
    \lim_{N\rightarrow \infty}
    \frac{1}{N}\sum_{n=0}^{N-1}\sum_{m=0}^{N-1}
    \left\{
    \frac{\partial V}{\partial \theta_0}\big(I,\theta_0+\Omega(I)n\big)
    \times \right. \nonumber \\
    &\left. \frac{\partial V}{\partial \theta_0}\big(I,\theta_0+\Omega(I)m\big)
    E[\xi_n\xi_m ]
    \right\},
  \end{IEEEeqnarray}
\end{itemize}
where
\begin{equation*}
  \frac{\partial V}{\partial \theta_0}\big(I,\theta_0+\Omega(I)n\big)
  \equiv
  \frac{\partial V}{\partial \theta}\big(I,\theta\big)_{\arrowvert \theta=\theta_0+\Omega(I)n}.
\end{equation*}

The second assumption is a condition on the losing memory rate of the process $\xi_n$
and the function $\varphi(n)$ is a measure  of the independence between past and future.
This condition is necessary if one wants to approximate the action dynamics with a diffusion
process.
The third requisite avoids the appearance of resonances between the unperturbed motion
and the deterministic component of the perturbation, $V(I,\theta)$.
Due to this, the diffusion coefficient turns out to be independent of the initial angle ($\theta_0$).

Being fulfilled these conditions, the following thesis is valid.
Introducing the {\it slow time}  $L=\epsilon^2 n$, in the limit $\epsilon\rightarrow 0$
the stochastic process~$I_\epsilon(L)\equiv I(L/\epsilon^2)$ weakly converges in $\mathcal{R}$
towards a diffusion process $\hat{I}(L)$ whose PDF, $\rho(\hat{I},L)$, satisfies
a F-P equation of the form:
\begin{equation}
  \label{fokker-planck}
  \frac {\partial \rho } {\partial L } (\hat{I},L) =
  \frac {1}{2}
  \frac {\partial} {\partial \hat{I}}
  \left\{
  \mathcal{D}(\hat{I}) \frac{\partial \rho}{\partial \hat{I}} (\hat{I},L)
  \right\}.
\end{equation}
The concept of {\it weak convergence} must be understood as convergence
of the PDFs (from now on abbreviated as {\it distributions}) for a sequence of stochastic processes.
Let $\rho(I, 0)$ be the initial particle distribution and let $\rho_\epsilon(I, t)$
be its average evolution at a time $t$
according to the stochastic map~(\ref{stoch_pert_symp_map_aa})
for a finite value of $\epsilon$.
The following limit is valid:
\begin{equation}
  \label{weak_limit}
  \lim_{\epsilon\rightarrow 0}\rho_\epsilon(I,L/\epsilon^2) = \rho(I, L),
\end{equation}
where $\rho(I,L)$ is the solution of the FP equation~(\ref{fokker-planck}).
This limit is the definition of convergence of the distribution.
Therefore Eq.~(\ref{fokker-planck}) can be used to get an approximation
of the true distribution function.
Under this point of view, $\mathcal{D}(I)$ is a rescaled diffusion coefficient.
Taking into account the Fourier cut--off, Ref.~\cite{1998Bazzani_proceeding}
proves  that the {\it analytical diffusion coefficient}, given by:
\begin{equation}
  \label{dif_coef_analytical}
  \mathcal{D}_a(I)\equiv \sum_{k}^{|k|\le k_{max}}k^2|V_k(I)|^2\tilde{\phi}(k\Omega(I)),
\end{equation}
is valid up to $\mathcal{O}(\epsilon)$,
where the spectral density of the noise:
\begin{equation}
  \label{spectral_dens}
  \tilde{\phi}(\nu) = \sum_{j=-\infty}^{\infty} \phi(|j|)\mathrm{e}^{\mathrm{i} j \nu}.
\end{equation}
and the autocorrelation function:
  \begin{equation*}
    \phi(n,m)\equiv E[\xi_n\xi_m]-E[\xi_n]E[\xi_m]= E[\xi_n\xi_m],
  \end{equation*}
have been introduced.
Due to the fact that the noise is stationary,
we have that $\phi(n,m)= \phi(n-m,0)\quad\forall n,m\in\mathbb{N}$,
so that in some opportunities we will use the notation
$\phi(n)\equiv \phi(n,0)$.

Ref.~\cite{1998JPhA...31.5843B} remarks the importance of the spectral
density in the behavior of the diffusion coefficient.
$\tilde{\phi}(\nu)$ contributes to $\mathcal{D}_a(I)$ only with the amplitudes
of the frequencies $\nu=k\Omega(I)$ which enter in the Fourier
expansion~(\ref{Fourier_pot}) of the perturbation. As a consequence,
the diffusion is enhanced when the spectral density of the noise is peaked
at the frequencies $k\Omega(I)$.

Let us explain this issue considering a simple situation in which
the spectral density is zero except in a neighborhood of radius $\Delta \nu$
of a fixed value $\nu_0$:
\begin{equation*}
  \tilde{\phi}(\nu) =
  \begin{cases}
    1 & \text{if } |\nu-\nu_0|<   \Delta \nu,\\
    0 & \text{if } |\nu-\nu_0|\geq\Delta \nu.
  \end{cases}
\end{equation*}
Thus, $\mathcal{D}_a(I)$ will be different from zero only in those actions
that satisfy simultaneously that $\nu_0-\Delta \nu <k\Omega(I)< \nu_0+\Delta \nu$ and
$V_k(I)\neq0$.
If we also assume that $\Omega$ has an inverse function, denoted by $\Omega^{-1}$,
it can be said that $\mathcal{D}_a(I)$ will be different from zero around
the action values $I=\Omega^{-1}(\nu_0/k)$.

\section{\label{sec:Gaussian_example} An example with a free rotator perturbed by colored Gaussian noise}
\noindent
In order to show how this approach can be applied, we consider the stochastic
symplectic map:
\begin{align}
  \label{map}
  I_{n+1}&= I_n - \epsilon \frac{\partial H_1}{\partial \theta_n}(I_{n+1}, \theta_n, \xi_n) \nonumber \\
  \theta_{n+1}&=\theta_n + \Omega(I_{n+1}) = \theta_n + I_{n+1} \hspace{1cm} mod \hspace{0.1cm} 2\pi
\end{align}
where
\begin{equation}
  \label{H_1}
  H_1(I, \theta, \xi)= H_1(\theta,\xi)=\xi V(\theta)=\xi \cos\theta.
\end{equation}
In other words, our system is a free rotator, with angle $\theta$ and angular velocity
$\Omega(I)\equiv I$, perturbed with a stochastic term.

As the Fourier series of our potential is:
\begin{equation}
  \label{potential}
  V(\theta) = \cos (\theta) = \frac{1}{2}\mathrm{e}^{\mathrm{i}\theta} +\frac{1}{2}\mathrm{e}^{-\mathrm{i}\theta},
\end{equation}
we have that
\begin{equation*}
  V_k(I) =
  \begin{cases}
    \frac{1}{2}  & \text{if } k=\pm 1,\\
    0             & \text{if } k\neq \pm 1.
  \end{cases}
\end{equation*}
The existence of only two terms in Eq.~(\ref{potential})
implies that it is not necessary to adopt a cut--off value.
Besides, together with the third hypothesis and the fact that
$\Omega(I)= I$, we have that:
\begin{equation}
  \label{U_region}
  \mathcal{R} =\{ I \in \mathbb{R} :\quad I \neq  2\pi q, \quad q\in\mathbb{Z} \}.
\end{equation}

Thus, the analytical expression~(\ref{dif_coef_analytical}) takes the form:
\begin{equation}
  \label{dif_coef_analytic_potential_1}
  \mathcal{D}_a(I) = \frac{1}{4}\{ \tilde{\phi}(-I) + \tilde{\phi}(I) \}.
\end{equation}
This analytical result will be compared
with a numerically computed diffusion coefficient defined in terms
of the variance (for $N$ units of time):
\begin{equation}
  \label{dif_coef_numerical}
  \mathcal{D}_{nu}(I;N)\equiv\frac{\langle (I_{N}-\langle I_{N}\rangle)^2\rangle } {\epsilon^2 N},
\end{equation}
where $\langle \cdot \rangle$ denotes a numerical average on the noise realization.
For a finite number of noise realizations, $N_r$, the numerical average
of an arbitrary quantity $A$ at time $t$ is given by:
\begin{equation*}
  \langle A_t \rangle\equiv \frac{1}{N_r}\sum_{k=1}^{N_r}A_t^{(k)},
\end{equation*}
where $A_t^{(k)}$ stands for the value associated to the $k^{th}$ realization.
Throughout this article we will use $N_r=10^5$.

We use a type of Gaussian colored noise which consists of an ensemble of
{\it damped stochastic harmonic oscillators}.
Following~\cite{1956Einstein_translation,1930PhRv...36..823U,
1943RvMP...15....1C,1945RvMP...17..323W,2011PhRvE..83d1103N},
the Langevin equation of a unit--mass damped Brownian particle
subject to the force field of a harmonic oscillator is:
\begin{equation}
  \label{langevin_1}
  \frac{d^2 \xi_t}{dt^2}=-\lambda \frac{d\xi_t}{dt}- \omega^2 \xi_t + c^{1/2} \Gamma_t,
\end{equation}
where $\omega$, $\lambda$ and $c$ are positive constants to be
defined below and where $\Gamma_t$ denotes a normalized white--noise
process, satisfying:
\begin{equation*}
  E[\Gamma_t]=0,\quad E[\Gamma_t\Gamma_{t'}]=\delta(t-t');
  \qquad \forall t,t'.
\end{equation*}
The constant $\omega$ denotes the deterministic frequency of the unperturbed
($\lambda=c=0$) oscillator.
The constant $\lambda$ denotes the friction coefficient
whose inverse $\lambda^{-1}$ is a characteristic relaxation time.

For $\omega=\lambda=0$,
$d\xi/dt$ is a Wiener process
with diffusion coefficient $c$. Besides, this constant
is related to the asymptotic diffusion coefficient, $D_B$,
that the Brownian particle would have
in case of null potential ($\omega=0$), like this~\cite{1996AmJPh..64..225G}:
\begin{equation*}
  D_B=c/2\lambda^2.
\end{equation*}

%
%
Eq.~(\ref{langevin_1}) can be explicitly solved~\cite{1930PhRv...36..823U,1943RvMP...15....1C}
and assuming $\omega^2-\lambda^2/4>0$, the analytic solution is:
\begin{IEEEeqnarray}{ll}
  \label{solution_damped_harm_osc}
  \xi_t =&~\frac{\lambda \xi_0+ 2 v_0}{2\omega_1}
  e^{-\frac{\lambda}{2}t} \sin(\omega_1t)
  +
  \xi_0  e^{-\frac{\lambda}{2}t} \cos(\omega_1t)
  \nonumber \\
  & +
  \frac{\sqrt{c}}{\omega_1}
  \int_0^t  e^{-\frac{\lambda}{2}(t-z)}\sin(\omega_1(t-z))dW_z,
\end{IEEEeqnarray}
where $\omega_1\equiv\sqrt{\omega^2-\lambda^2/4}$ is the single
proper frequency of the noise and where $dW_z$ is the differential of a Wiener process.

In this paper we consider deterministic initial distributions and each realization
corresponds to a realization of the Wiener process.
The process tends asymptotically, for $t\rightarrow +\infty$,
towards a stationary state with zero mean, with autocorrelation function
(see Appendix~\ref{app:autocorrel_damped_harm_osc}):
\begin{equation}
  \label{autocorrel_damped_harm_osc}
 \phi_s(\tau)=\frac{c}{2 \lambda \omega^2}e^{-\frac{\lambda}{2}\tau}
 \{\cos(\omega_1\tau)+\frac{\lambda}{2\omega_1}\sin(\omega_1\tau)\},
\end{equation}
and with variance:
\begin{equation*}
  \sigma_s^2=\phi_s(0)=\frac{c}{2 \lambda \omega^2}.
\end{equation*}
%

%
%
We have performed a numerical stochastic integration
for parameter values given by $\omega=1$, $\lambda=0.2$ and $c=0.4$ and for
an initial condition given by $(\xi_0,v_0)=(-3,0)$.
This corresponds to $\omega_1 \approx 0.995$.
We have used the
{\it sderk}\footnote{Written by Daniel Steck, {\it http://steck.us/computer.html}.}
integration package. It contains many schemes for obtaining strong
solutions to stochastic differential equations. We have used
one which consists of a fourth order Runge-Kutta scheme for the deterministic part
and a first order for the stochastic part.
We considered an ensemble of $N_r$ noise realizations during
a total integration time of $t_T=10^3$.
We have numerically estimated a relaxation time $t_s=500$ such that for $t>t_s$ the system can be considered in a stationary state.

\begin{figure}[ht!]
  \includegraphics[width=0.49\textwidth]{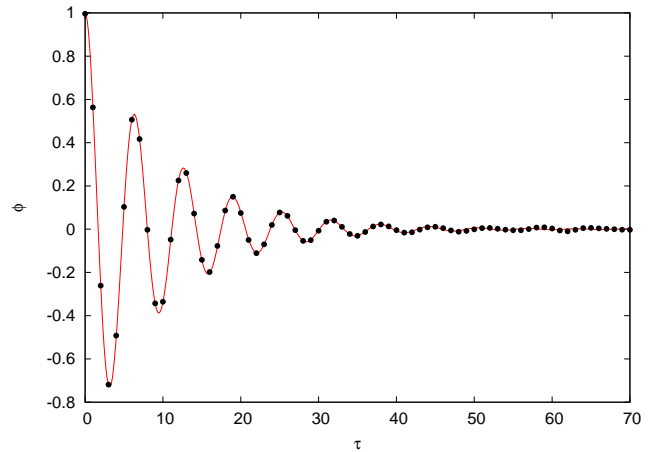}
  \caption{\label{fig:autocorrel_damped_osc}
    (Color online)
    Autocorrelation function of an ensemble of damped stochastic harmonic oscillators  for parameter values
    $\omega=1$, $\lambda=0.2$ and $c=0.4$ and for an initial condition given by $(\xi_0,v_0)=(-3,0)$.
    The asymptotic analytical function given in Eq.~(\ref{autocorrel_damped_harm_osc}) is displayed with a red solid curve. The numerical function, for $N_r=10^5$ noise realizations and starting to measure
    since $t_s=500$, is displayed with black dots.}
\end{figure}

Thus, in Fig.~\ref{fig:autocorrel_damped_osc}, we display
$\phi(t_s,t_s+\tau)$, for $0 \leq\tau\leq 70$, with black dots.
In the same figure, the analytic solution for $\phi_s(\tau)$ is illustrated with
a red solid curve.

\begin{figure}[ht!]
  \includegraphics[width=0.49\textwidth]{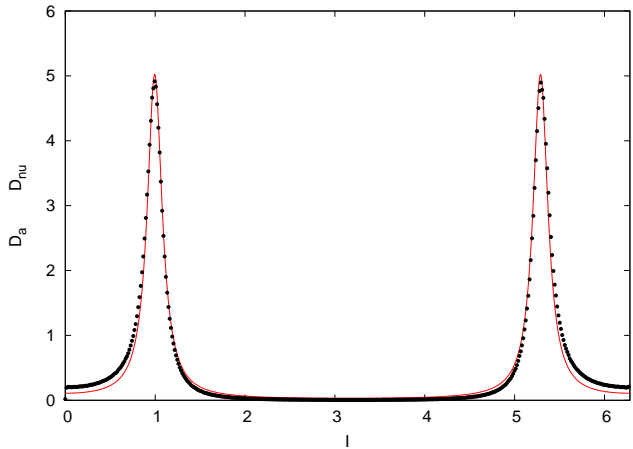}
  \caption{\label{fig:dif_coef_damped_harm_osc}
    (Color online)
    $\mathcal{D}_a(I)$ (red solid curve)
    and $\mathcal{D}_{nu}(I;500)$ (black dots)
    for the map given by Eq.~(\ref{map}).
    As noise we have used a stationary ensemble of damped harmonic oscillators
    with parameter values $\omega=1$, $\lambda=0.2$ and $c=0.4$ and with an initial
    condition given by $(\xi_0,v_0)=(-3,0)$.
    In the case of the numerical coefficient we have used
    $\epsilon=10^{-7}$, an initial angle given by $\theta_0\approx 0.214$
    and the values of the same noise realizations computed
    for the previous figure, taking into account only the
    time interval: $t_s\leq t \leq 2 t_s$ ($t_s=500$).}
\end{figure}

The formula for the
spectral density associated to this asymptotic state
is given by Eqs.~(\ref{spect_dens_damped_harm_osc_1}),~(\ref{spect_dens_damped_harm_osc_2})
and~(\ref{spect_dens_damped_harm_osc_3}).
Introducing it into Eq.~(\ref{dif_coef_analytic_potential_1}) we have
computed  $\mathcal{D}_a(I)$ and displayed it in
Fig.~\ref{fig:dif_coef_damped_harm_osc} with a red solid curve.
In the same figure, we display with black dots, the quantity
$\mathcal{D}_{nu}(I;500)$, for 500 values of $I$ placed
equidistantly in the interval $[0,2\pi)$,
where we have used  $\epsilon=10^{-7}$, an initial angle $\theta_0\approx 0.214$
and the values of the same noise realizations computed
for the previous figure, taking into account only the
time interval: $t_s\leq t \leq 2 t_s$.
All over this article, we keep fixed the value $\theta_0$, unless
explicitly stated otherwise.
We can see that both coefficients agree for every action value
and that the biggest difference takes
place in the neighborhood of $I=0$ and $I=2\pi$, which are points that
are outside of the region $\mathcal{R}$ defined by Eq.~(\ref{U_region}).
Moreover, we see that the diffusion coefficient reaches its maximum
value for $I=\omega_1$ and $I=2\pi-\omega_1$. This agrees with
the analytical condition for diffusion enhancement that applies
to this system: $\omega_1=\pm \Omega(I)=\pm I$ (mod $2\pi$).

We have carried out other numerical experiments changing $\epsilon$
and the result is similar as long as $\epsilon \lesssim 10^{-4}$.

Now we point out a connection between the
damped harmonic stochastic oscillator and the stochastic rotator.
On the one hand we have that, for the chosen
parameter values of the harmonic noise, the contribution of
the term with the {\it sine} function in $\phi_s$
is quite small, in such a way that neglecting this term
hardly produces any change in $\mathcal{D}_a$.
On the other hand we know~\cite{1998JPhA...31.5843B} that the
exact analytical asymptotic autocorrelation function of the
stochastic rotator is of the form $\phi_{sr}(\tau)=a e^{-b\tau}\cos(\omega_\star \tau)$,
for some parameters $a,b$ and $\omega_\star$.
Thus, we conclude that the diffusion coefficient produced by a coupling with
a stochastic rotator is similar to the one produced by a coupling
with the harmonic noise whenever the condition $\lambda << 2\omega_1$ is satisfied.

In the next section we will discuss the stochastic approach to the
computation of the diffusion coefficient in chaotic systems.

\section{\label{sec:method} A semi--numerical method for chaotic systems}
\noindent
In the previous sections we have seen that the autocorrelation function
of the noise has a net effect in the diffusion.
In fact, the expression for $\mathcal{D}(I)$ depends
on the trajectories of the integrable part and on the autocorrelation
function of the noise and it is related with the
Taylor-Green-Kubo formula (see~\cite{2011PhRvE..83d6402S} and references therein).

We wonder whether the hypothesis of the averaging theorem, whose result
is the F-P Eq.~(\ref{fokker-planck}) with diffusion coefficient
given by Eq.~(\ref{dif_coef_analytical}), could be modified
in order to admit deterministic perturbations.
We will show empirically that a particular chaotic perturbation can produce a diffusion
process in the integrable part that can be modelled by the mentioned kinetic equation
and will give a semi--numerical method to estimate the diffusion coefficient.

In order to show how we intend to apply the stochastic formalism to
a symplectic map with divided phase space, we will introduce two intermediate
deterministic systems that link the one given by Eqs.~(\ref{map}) and~(\ref{H_1})
with the one given by Eq.~(\ref{map_Meiss}), being the latter the main
object of study of this article.

The first intermediate map is built by
replacing, in Eq.~(\ref{H_1}), the stochastic process, $\xi$, by
an ensemble of chaotic trajectories.
In particular, we set $\xi_n=\sin(\psi_n)$, where $\psi_n$ is the angle
at time $n$ that corresponds to a chaotic trajectory of the SM.
Thus, we have a (pseudo) sample space, $\mathcal{S}$, which is some subset
of the trajectories that belong to a particular chaotic layer of the SM.
Due to the unicity of the solution of initial value problems for
deterministic systems, it is possible to label $\mathcal{S}$
unambiguously with the values of the initial conditions in the $[J,\psi]$ plane.
In Secs.~\ref{sec:thick_chaotic_layer} and~\ref{sec:thin_chaotic_layer} we will give explicit
expressions for the sample spaces.
Thus, we have the following skew coupled map:
\begin{align*}
  I_{n+1}&= I_n - \epsilon  \sin(\psi_n) \sin(\theta_n) \nonumber \\
  \theta_{n+1}&= \theta_n + I_{n+1} \hspace{2cm} mod \hspace{0.1cm} 2\pi \nonumber \\
  J_{n+1}&= J_n + K \sin(\psi_n) \nonumber \\
  \psi_{n+1}&= \psi_n + J_{n+1}  \hspace{2cm} mod \hspace{0.1cm} 2\pi.
  \label{skew_map_1}
\end{align*}

The second intermediate map is constructed
by gene-ralizing the perturbation given in Eq.~(\ref{H_1}) to:
\begin{equation*}
  H_1(\theta, \psi)=\cos(\theta+\psi),
\end{equation*}
obtaining the following (also partially coupled) map:
\begin{align}
  I_{n+1}&= I_n - \epsilon \sin(\theta_n + \psi_n ) \nonumber \\
  \theta_{n+1}&= \theta_n + I_{n+1} \hspace{2cm} mod \hspace{0.1cm} 2\pi \nonumber \\
  J_{n+1}&= J_n + K \sin(\psi_n) \nonumber \\
  \psi_{n+1}&= \psi_n + J_{n+1}  \hspace{2cm} mod \hspace{0.1cm} 2\pi.
  \label{skew_map_2}
\end{align}
The statistical properties of hyperbolic maps are stable under
small enough perturbations. Although the variables $(J,\psi)$
do not perform an hyperbolic dynamics, they have approximately
this quality in any chaotic component of the SM.

We remark that the difference between map~(\ref{skew_map_2}) and the symplectic
map~(\ref{map_Meiss}) is the presence of feedback coupling between the slow dynamics and the
chaotic dynamics $(J,\phi)$. In the case of strong chaos, $K\gg 1$, a thick chaotic layer exists in the phase space and
we do expect that the $\mathcal{O}(\epsilon)$ perturbation
 would not affect the diffusion
process associated to the rotator action $I$. Conversely in the case of weak chaos, $K\simeq 1$, the presence of partial
barriers in the phase space $(J,\psi)$ could be affected by a small perturbation even in the diffusion limit, and
the statistical properties of the ``noise'' $\psi_n$ could depend on the long time evolution so that the assumption of
stationary noise cannot be applied. However we have numerically checked that the effect of this back coupling is negligible when computing
diffusion coefficients for the slow variables $(I,\theta)$.

Thus, our idea is to model diffusion in the symplectic map (\ref{map_Meiss}) by applying
a semi--numerical scheme (related to the averaging theorem) to map~(\ref{skew_map_2}).
It can be demonstrated, following closely the proof in~\cite{1998Bazzani_proceeding},
that for a general perturbation of the form:
\begin{equation*}
  H_1(I, \theta, \psi ) = \sum_k h_k(I) \mathrm{e}^{\mathrm{i} k(\theta + \psi)},
\end{equation*}
the analytic diffusion coefficient is given by:
\begin{equation*}
  \mathcal{D}_a(I)= \sum_k^{|k|\le k_{max}} k^2 h_k^2(I)\tilde{\phi}_{k,k}(k\Omega(I))
  , \hspace{1.0cm} \forall \hspace{0.5mm} I\in \mathcal{R},
\end{equation*}
where $\tilde{\phi}_{k,k}(\nu)$ are the spectral densities associated to the
following autocorrelation functions of the SM:
$\phi_{k,k'}(n,m)= E[\mathrm{e}^{\mathrm{i}k\psi_n} \mathrm{e}^{\mathrm{i}k'\psi_m}]
- E[\mathrm{e}^{\mathrm{i}k\psi_n}]E[ \mathrm{e}^{\mathrm{i}k'\psi_m}]$.
As there are no analytical formulae for this autocorrelation
functions valid for sufficiently long times, i.e. times of the order of $500$,
we will develop a semi--numerical formula.

The generalization of the
expression of $\mathcal{D}(I)$ given in Eq.~(\ref{lim_coef_dif}),
in the case of map~(\ref{skew_map_2}), is:
\begin{IEEEeqnarray}{ll}
  \label{lim_coef_dif_generalized}
  \mathcal{D}(I)\equiv &
  \lim_{N\rightarrow \infty}
  \frac{1}{N}\sum_{n=0}^{N-1}\sum_{m=0}^{N-1}
  \Big\{
  E\Big[ \frac{\partial H_1}{\partial \theta_0}\big(\theta_0+\Omega(I)n,\psi_n \big)
    \times  \nonumber \\
    &\frac{\partial H_1}{\partial \theta_0}\big(\theta_0+\Omega(I)m,\psi_m \big)
  \Big] \nonumber \\
  & - E\Big[\frac{\partial H_1}{\partial \theta_0}\big(\theta_0+\Omega(I)n,\psi_n \big)\Big] 
  \times \nonumber \\
  & E\Big[\frac{\partial H_1}{\partial \theta_0}\big(\theta_0+\Omega(I)m,\psi_m \big)\Big]
  \Big\}.
\end{IEEEeqnarray}
If instead of taking the limit, we evaluate numerically at a finite time $N$,
we have the {\it semi--numerical diffusion coefficient}:
\begin{IEEEeqnarray}{l}
  \label{dif_coef_symplectic_b}
  \mathcal{D}_{sn}(I;N) \equiv \nonumber \\
  \frac{1}{N}\sum_{n=0}^{N-1}\sum_{m=0}^{N-1}
  \Big\{ \phi_1(n,m)  \cos\big(\theta_0+\Omega(I)n\big)\cos\big(\theta_0+\Omega(I)m \big)  \nonumber \\
  \quad + \phi_2(n,m) \sin\big(\theta_0+\Omega(I)n\big)\sin\big(\theta_0+\Omega(I)m \big)  \nonumber \\
  \quad + \phi_3(n,m) \cos\big(\theta_0+\Omega(I)n\big)\sin\big(\theta_0+\Omega(I)m \big)  \nonumber \\
  \quad + \phi_4(n,m) \sin\big(\theta_0+\Omega(I)n\big)\cos\big(\theta_0+\Omega(I)m \big) \Big\};
\end{IEEEeqnarray}
where now the autocorrelation functions $\phi_i$ ($i=1,\dots,4$) should be
computed numerically:
\begin{align}
  \label{phi_i_esperanza}
  \phi_1(n,m)&=  \langle \sin\psi_n \sin\psi_m \rangle
  - \langle \sin\psi_n\rangle  \langle\sin\psi_m \rangle,\nonumber   \\
  \phi_2(n,m)&=  \langle \cos\psi_n \cos\psi_m \rangle
  - \langle \cos\psi_n \rangle  \langle\cos\psi_m \rangle,\nonumber  \\
  \phi_3(n,m)&=  \langle \sin\psi_n \cos\psi_m \rangle
  -\langle \sin\psi_n \rangle  \langle\cos\psi_m \rangle,\nonumber  \\
  \phi_4(n,m)&=  \langle \cos\psi_n \sin\psi_m \rangle
  - \langle \cos\psi_n \rangle  \langle\sin\psi_m \rangle.
\end{align}
For sufficiently large values of $N$, $\mathcal{D}_{sn}(I;N)$ is effectively independent of $\theta_0$.

\section{\label{sec:thick_chaotic_layer} Thick chaotic layer }
\noindent
In this section we will test the stochastic approach to describe the diffusion in the 
symplectic map~(\ref{map_Meiss}), using a parameter $K=3$.

For this value, the measure of initial conditions in the phase
space that give rise to chaotic orbits is much larger than the measure associated
to regular orbits. The area of the islands of stability is
small and almost every chaotic orbit belongs to the 
thick layer, also called {\it chaotic sea}.
Moreover, the $J$ variable is unbounded for some trajectories
so that the chaotic sea has an infinite extent.
Notwithstanding, due to the $2\pi$--periodicity of the perturbation
functions, it is plausible to compactify $J$ to the interval
$[0,2\pi)$.

In order to compute $\mathcal{D}_{sn}(I,N)$, the first step is to compute
the numerical autocorrelation
functions $\phi_i$ defined in the previous section.
Similarly to what was done in Sec.~\ref{sec:Gaussian_example},
we select a time, $n_s=500$, above which, the ensemble
of SM orbits are close to a stationary state (in the compactified
phase space).
This choice allows the ensemble to distribute all over the chaotic sea.
As sample space we choose an ensemble of $N_r$ trajectories of the SM,
with {\it seeds} placed along the straight line $J=J_0\equiv3$ and considering
only the time interval: $n_s \leq n \leq 2 n_s$. Let $\mathcal{S}_n$ be the ensemble of the sampled
orbits at time $n$:
\begin{IEEEeqnarray}{ll}
  \label{IC_sample_space_K3.0}
  \mathcal{S}_{n}\equiv \{& (J_{n}^{(k)},\psi_{n}^{(k)})=S^{n}[(J_0, u_k )]:\nonumber \\
  & u_k\in \mathcal{U}(0,2\pi);\nonumber \\
  & k=1,2,\dots,N_r;\quad N_r=10^5 \},
\end{IEEEeqnarray}
where $S^n[\cdot]$ denotes $n$ iterations of the SM
and $\mathcal{U}(a,b)$ denotes the
uniform distribution in the segment $(a,b)$,
we define $\mathcal{S}$ as the union of $\mathcal{S}_n$
with $n=n_s,n_s+1,\ldots,2n_s$ and $n_s=500$.

\begin{figure}[ht!]
  \includegraphics[width=0.49\textwidth]{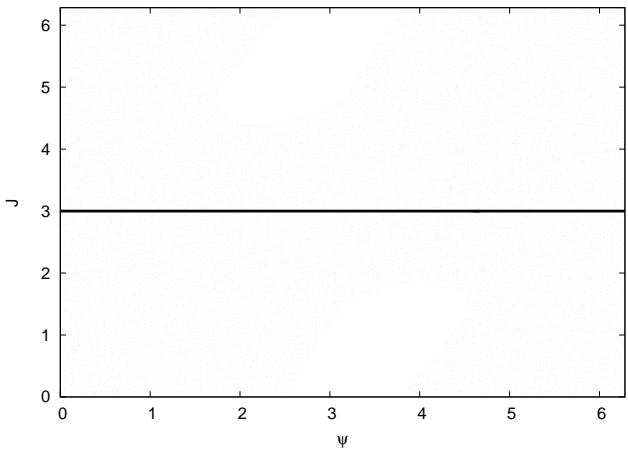}
  \caption{Snapshots of the sample space associated to the SM
    with $K=3$: $\mathcal{S}_{0}$ is displayed with a black straight line
    while $\mathcal{S}_{n_s}$ ($J$ compactified to the interval $[0,2\pi)$)
    is displayed with dots.
    \label{fig:SM_K3.0_snapshot_sample_space}
    }
\end{figure}

In Fig.~\ref{fig:SM_K3.0_snapshot_sample_space} we can see the
ensemble at both its initial state (seeds), $\mathcal{S}_{0}$, displayed
with a black straight line, and at its nearly--stationary state,
$\mathcal{S}_{n_s}$, displayed with dots.

\begin{figure}[ht!]
  \begin{center}
    \begin{tabular}{c}
      {\includegraphics[width=0.48\textwidth]{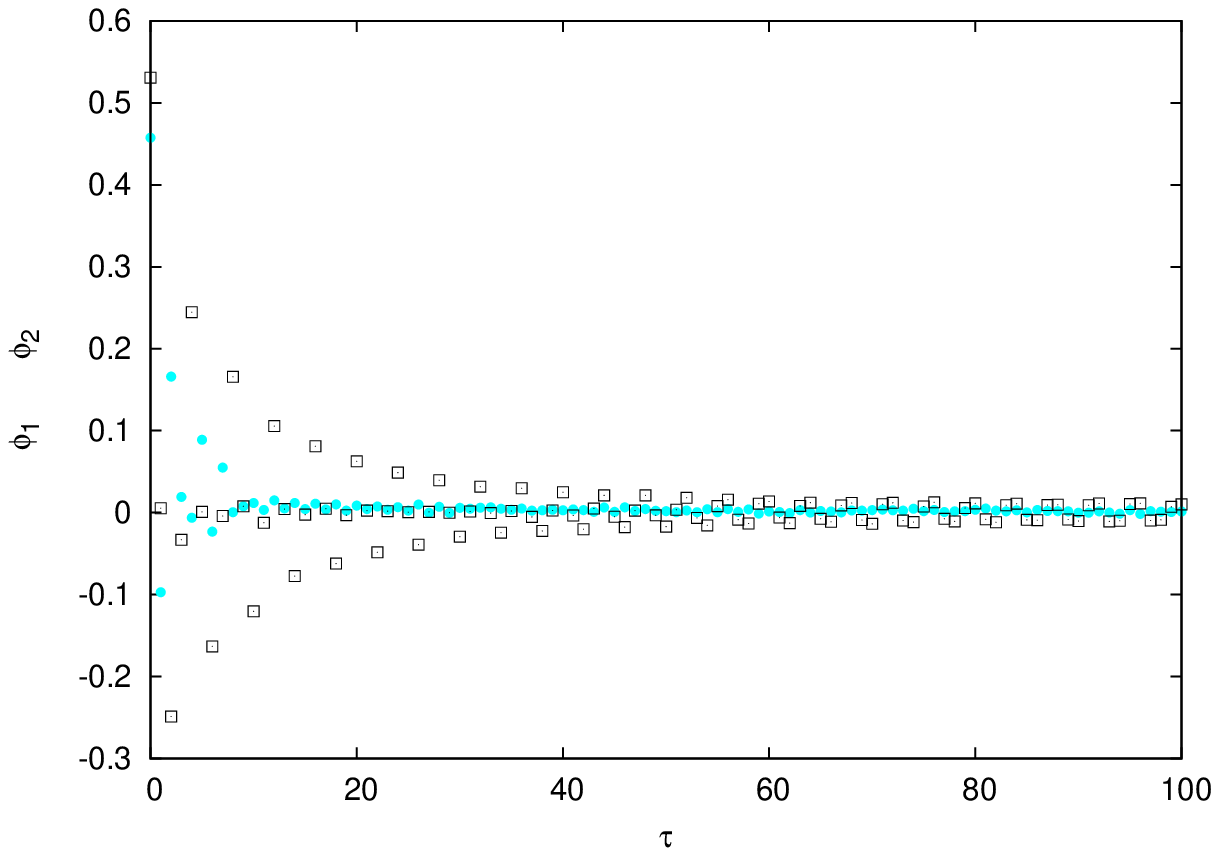}}\\
      {\includegraphics[width=0.49\textwidth]{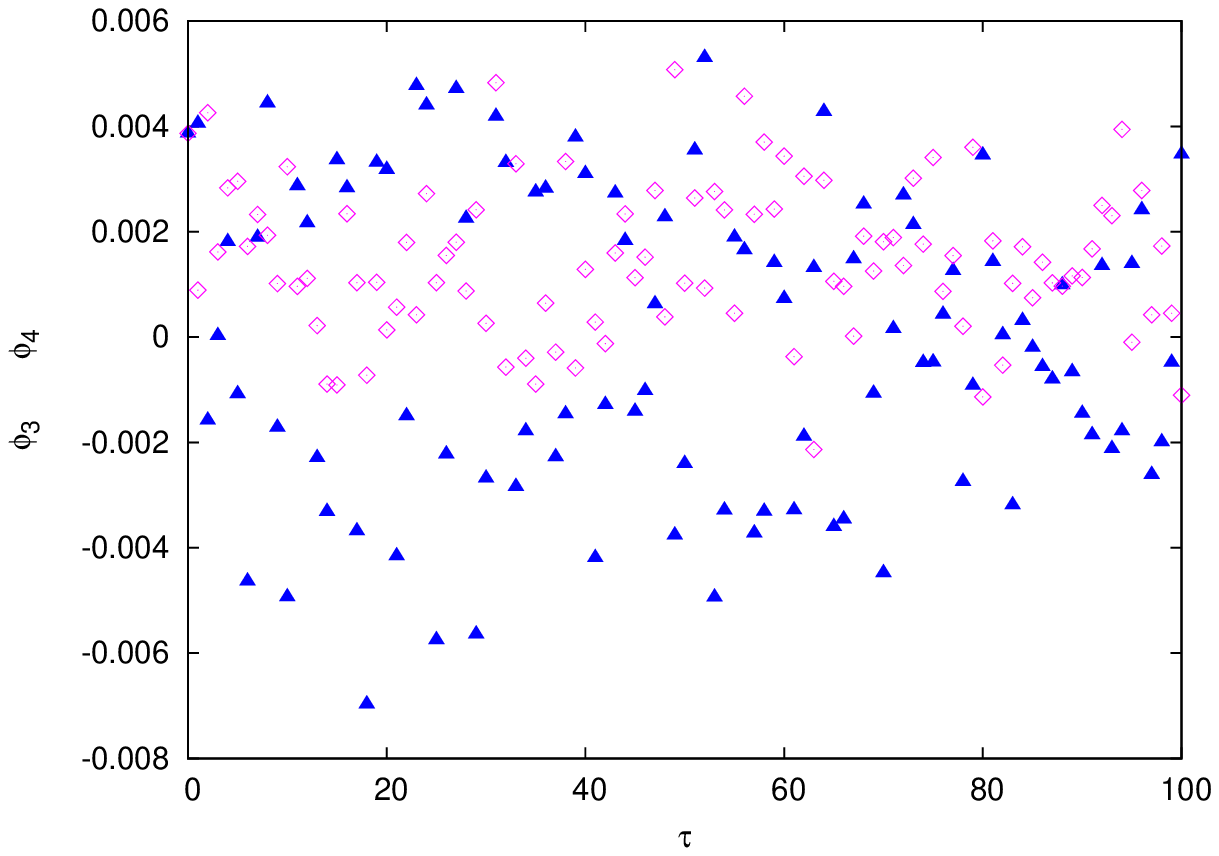}}
    \end{tabular}
    \caption{(Color online) SM autocorrelation 
      functions: $\phi_i(n_s,n_s+\tau)$ ($n_s=500$) of
      Eq.~(\ref{phi_i_esperanza}), for $i=1,2,3,4$
      using black (square) dots, cyan (round) dots,
      blue (triangle--shaped) dots and magenta (diamond--shaped) dots, respectively.
      The sample set is $\mathcal{S}=\bigcup_{n=n_s}^{2n_s}\mathcal{S}_{n}$,
      with $\mathcal{S}_{n}$ defined by Eq.~(\ref{IC_sample_space_K3.0}).
      \label{fig:SM_K3.0_phi_1234}
    }
  \end{center}
\end{figure}

In Fig.~\ref{fig:SM_K3.0_phi_1234} we display, for  $0\leq\tau\leq 100$, the values of
$\phi_i(n_s,n_s+\tau)$, for $i=1,2,3,4$ using
black (square) dots, cyan (round) dots,
blue (triangle--shaped) dots and magenta (diamond--shaped) dots, respectively.

We notice that all of them oscillate in a neighborhood of zero,
that $\phi_1$ presents oscillations of higher amplitude than the other cases
and that $\phi_3$ and $\phi_4$ have relatively small values at every moment.
Besides, we appreciate that the autocorrelations do not decay
to zero for  $\tau>>1$, unlike what happened in the  colored Gaussian case.
In spite of this difference, there is a
kind of similarity between $\phi_s$ and $\phi_1$ because
a least square fit of the ansatz
$\phi_1(\tau)=(a_0 + a e^{-b\tau})\cos(\omega_\star \tau)$
(for parameters $a_0,a,b$ and $\omega_\star$) worked very well,
even up to times of the order of $\tau\sim 300$,
giving a proper frequency $\omega_\star \approx1.572\approx\pi/2$.

\begin{figure}[ht!]
  \begin{center}
    \begin{tabular}{c}
      {\includegraphics[width=0.49\textwidth]{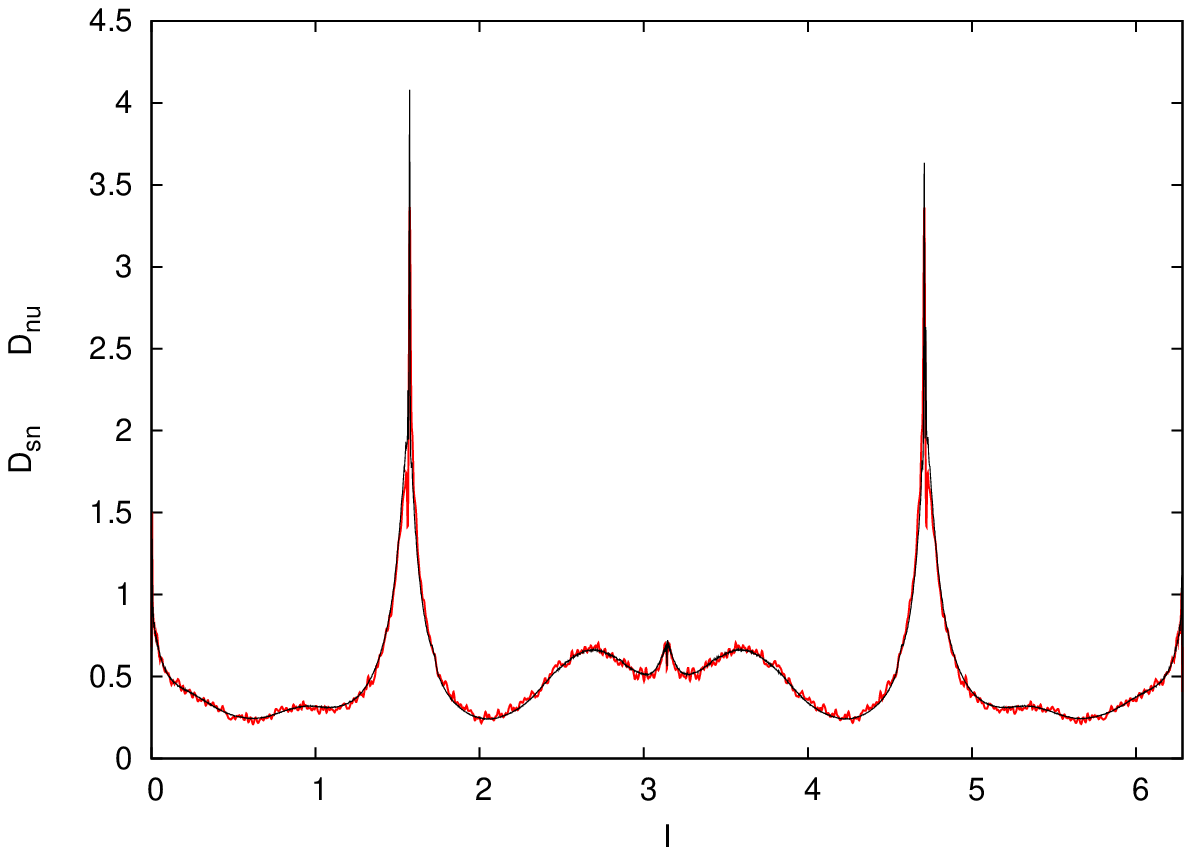}}\\
      {\includegraphics[width=0.49\textwidth]{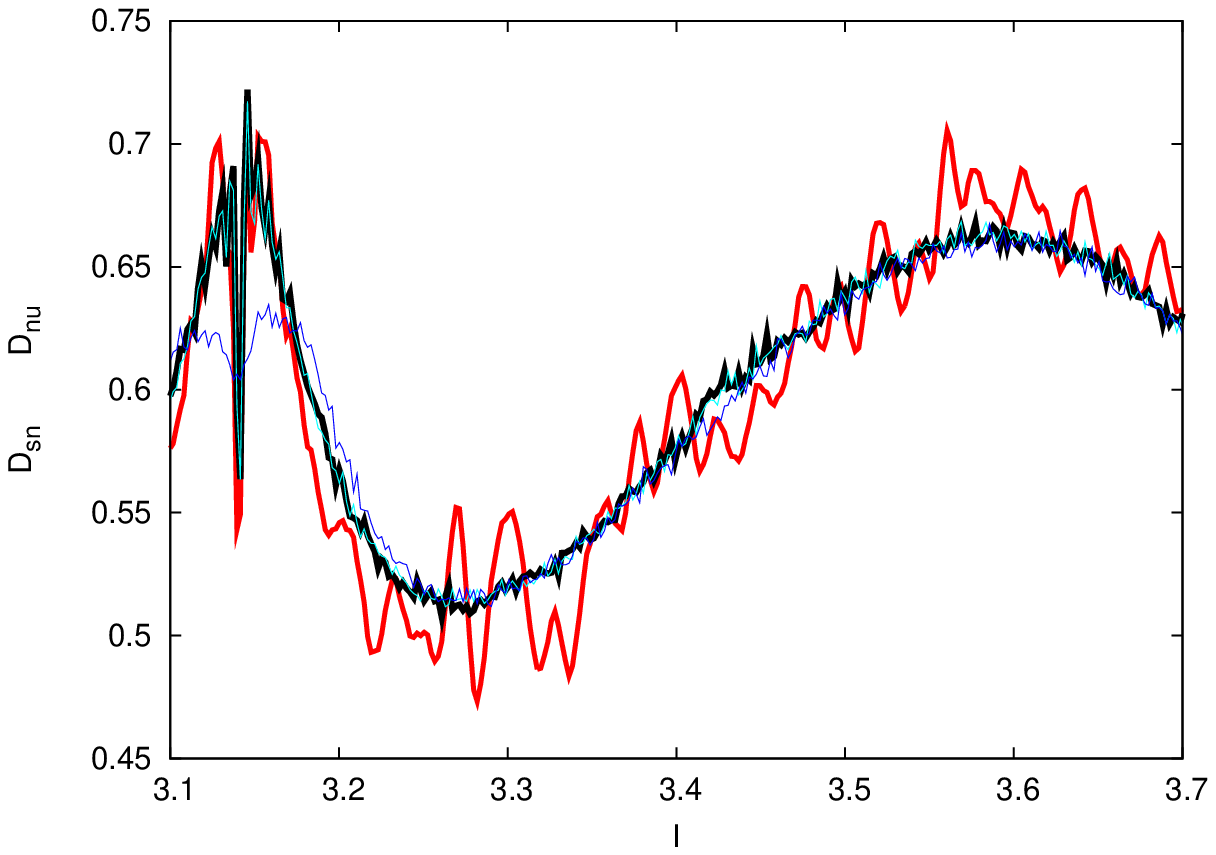}}
    \end{tabular}
    \caption{(Color online)
      Diffusion coefficients for the symplectic map (\ref{map_Meiss}),
      using $K=3$ and $\mathcal{S}=\bigcup_{n=n_s}^{2n_s}\mathcal{S}_{n}$
      with $\mathcal{S}_{n}$ defined by Eq.~(\ref{IC_sample_space_K3.0}).
      $\mathcal{D}_{sn}(I;500)$ is shown in red.
      $\mathcal{D}_{nu}(I;500)$ was computed with $\epsilon=10^{-3}, 10^{-5}, 10^{-7}$
      and displayed in colors blue, cyan and black, respectively.
      \label{fig:prediction_coef_K3.0}
    }
  \end{center}
\end{figure}

We have introduced the previously computed four autocorrelation functions
into Eq.~(\ref{dif_coef_symplectic_b}) in order to calculate
$\mathcal{D}_{sn}(I;500)$, for $3000$ equidistant values of $I$ in the interval $[0,2\pi)$.
In Fig.~\ref{fig:prediction_coef_K3.0}--top
we compare this value with
the numerical diffusion coefficient associated to the symplectic map (\ref{map_Meiss})
for $\epsilon=10^{-7}$, in red and black colors, respectively.
For each value of $I$, the initial conditions of the ensemble chosen
to compute $\mathcal{D}_{nu}(I,500)$, according to Eq.~(\ref{dif_coef_numerical}),
consist of the direct product between the point $(I,\theta_0)$ and
$\mathcal{S}_{n_s}$.

The result is that $\mathcal{D}_{sn}$ succeeds in predicting $\mathcal{D}_{nu}$.
In Fig.~\ref{fig:prediction_coef_K3.0}--bottom we display, for $3.1\leq I\leq3.7$,
the same two curves of the top panel together with
$\mathcal{D}_{nu}(I;500)$ computed with $\epsilon=10^{-3}$ and $\epsilon=10^{-5}$
in colors blue and cyan, respectively.

We observe that the numerical coefficients for $\epsilon=10^{-5}$ and $10^{-7}$
are equivalent to each other throughout the interval
whereas the one for $\epsilon=10^{-3}$ differs from the other
coefficients in the neighborhood of $I=\pi$.
This discrepancy is expected, because the method is meant to work
as long as $\epsilon$ is small enough, and
it is in agreement with the result commented
near the end of Sec.~\ref{sec:Gaussian_example}.

Comparing the numerical diffusion coefficients of Figs.~\ref{fig:dif_coef_damped_harm_osc}
and~\ref{fig:prediction_coef_K3.0}--top,
we see that the diffusion generated by coupling the free rotator with the SM
is different from the one generated by coupling the same integrable system with
an ensemble of damped stochastic harmonic oscillators.
The only aspect in common between both coefficients
is that they have absolute maximums in those actions whose associated frequency is
equal to any of the characteristic frequencies of the
perturbation (chaotic or stochastic).
If we had used an harmonic noise with proper frequency similar to
the one of the SM ($\omega_1 \approx \pi/2$), we would have obtained a result analogue to
the one of Fig.~\ref{fig:dif_coef_damped_harm_osc} but with its maximums at $I\approx \pi/2$
and $I\approx 3\pi/2$.
Taking this information into account together with the fact that
the diffusion driven by a stochastic rotator is similar to the one driven
by harmonic noise, we have given an example in which
the effect of coupling an integrable map to a chaotic perturbation
can not be modeled by the effect of coupling the same integrable map to a
stochastic rotator (neither to a damped stochastic harmonic oscillator).
Thus, we have refuted the conjecture cited in the Introduction.

Nevertheless, it  is true that a chaotic perturbation can drive
a process qualitatively and quantitatively similar to a diffusion process,
as it will become clear in the rest of this section, where we
will empirically show that the action, $I$,
of the 4D symplectic map (\ref{map_Meiss}) behaves as a diffusion process that satisfies
the F-P Eq.~(\ref{fokker-planck}).

We have numerically solved the F-P Eq.~(\ref{fokker-planck})
for an ensemble with initial conditions following a Gaussian distribution,
i.e. $I_0\sim\mathcal{N}(\mu,\sigma)$, with mean value
$\mu=1.75$ and standard deviation $\sigma=0.1$.
We used a Cranck--Nicholson implicit algorithm and
the values of $\mathcal{D}(\hat{I})$ where those of $\mathcal{D}_{nu}(I;500)$,
plotted in Fig.~\ref{fig:prediction_coef_K3.0}--top.
Using  as input data $\mathcal{D}_{sn}$, instead of $\mathcal{D}_{nu}$,
does not significantly change the results.

Besides, we have numerically computed the evolution of the symplectic map (\ref{map_Meiss}) of an ensemble
of $N_r$ test particles with initial conditions such that:

\begin{itemize}
\item $I_0\sim \mathcal{N}(1.75,0.1)$
\item $\theta_0=(1+\sqrt{5})/2\approx 1.618$
\item $(J_0,\psi_0)\in \mathcal{S}_{n_s}$.
\end{itemize}

We used $\epsilon=10^{-3}$ so that the relation between the
F-P time~($L$) and the (Hamiltonian) real time~($n$) is:
\begin {equation}
  \label{FP-symplectic_time_relation}
  L =\epsilon^2 n =10^{-6} n.
\end{equation}

\begin{figure}[ht!]
  \includegraphics[width=0.49\textwidth]{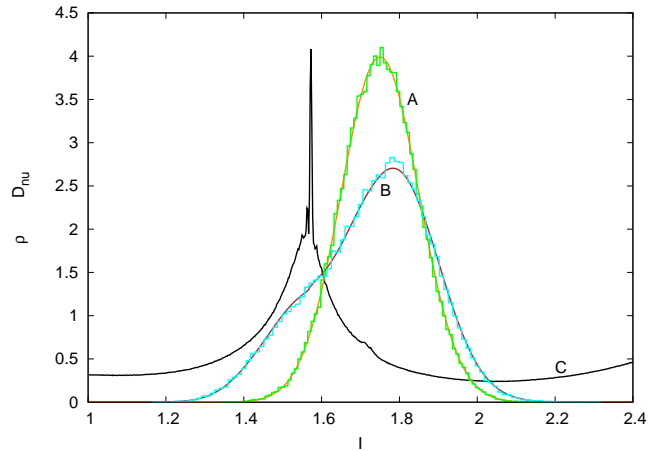}
  \caption{(Color online)
    Distributions for thick layer regime ($K=3$):
    $\rho(I,0)$, $\rho_{\epsilon}(I,0)$, $\rho(I,0.02)$ and $\rho_{\epsilon}(I,2\times10^4)$
    in colors orange (smooth line A), green (histogram A), brown (smooth line B)
    and cyan (histogram B), respectively.
    For the histograms the coupling parameter is $\epsilon=10^{-3}$.
    The black line (C) displays $\mathcal{D}_{nu}(I;500)$, the diffusion
    coefficient used when integrating the F-P equation.
    \label{fig:fokker_planck_K3.0_a}
  }
\end{figure}

\begin{figure}[ht!]
  \includegraphics[width=0.49\textwidth]{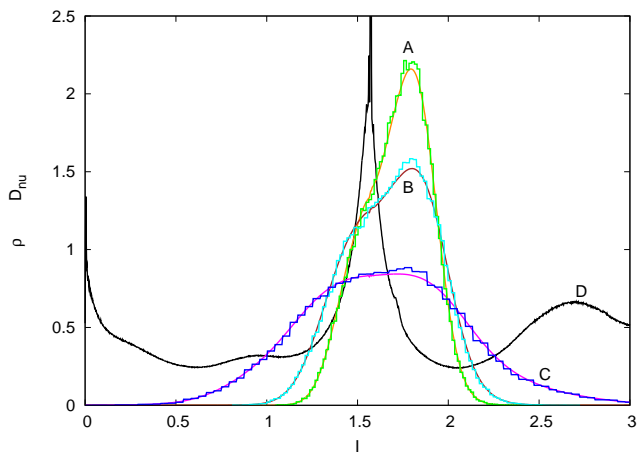}
  \caption{(Color online)
    Distributions:
    $\rho(I,0.04)$, $\rho_{\epsilon}(I,4\times 10^4)$,
    $\rho(I,0.10)$, $\rho_{\epsilon}(I, 10^5)$,
    $\rho(I,0.50)$ and $\rho_{\epsilon}(I,5\times 10^5)$
    in colors  orange (smooth line A), green (histogram A),
    brown (smooth line B), cyan (histogram B),
    magenta (smooth line C) and blue (histogram C), respectively.
    For the histograms the coupling parameter is $\epsilon=10^{-3}$.
    The black line (D) displays $\mathcal{D}_{nu}(I;500)$.
    \label{fig:fokker_planck_K3.0_b}
    }
\end{figure}

Figs.~\ref{fig:fokker_planck_K3.0_a} and~\ref{fig:fokker_planck_K3.0_b}
show both the F-P solution ($\rho$)
and histograms done with the orbits of the symplectic map ($\rho_{\epsilon}$).
In the former figure, we have plotted the functions
 $\rho(I,0)$, $\rho_{\epsilon}(I,0)$, $\rho(I,0.02)$ and $\rho_{\epsilon}(I,2\times10^4)$
in colors orange (smooth line A), green (histogram A), brown (smooth line B)
and cyan (histogram B), respectively.
It can be seen that the behavior of the action of the symplectic
map (\ref{map_Meiss}) resembles closely the diffusion process given by the F-P solution.
The diffusion coefficient is displayed in the same figure, in order to show
the reason why the left hand side of the initially Gaussian distribution diffuses
faster than its right hand counterpart.

The second figure shows
$\rho(I,0.04)$, $\rho_{\epsilon}(I,4\times 10^4)$,
$\rho(I,0.10)$, $\rho_{\epsilon}(I, 10^5)$,
$\rho(I,0.50)$ and $\rho_{\epsilon}(I,5\times 10^5)$
in colors orange (smooth line A), green (histogram A),
brown (smooth line B), cyan (histogram B),
magenta (smooth line C) and blue (histogram C), respectively.

\begin{figure}[ht!]
  \includegraphics[width=0.49\textwidth]{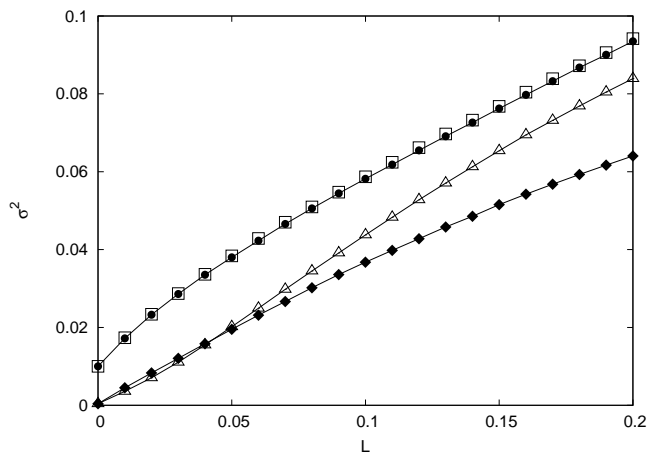}
  \caption{$\sigma^2(L)$, for $0 \leq L\leq 0.2$. The round dots correspond to $K=3$ and
    $\epsilon=10^{-3}$, the square dots correspond to $K=3$ and $\epsilon=10^{-4}$,
    the diamond--shaped dots correspond to $K=0.9$ and $\epsilon=10^{-3}$ and the
    triangle--shaped dots correspond to $K=0.9$ and $\epsilon=10^{-4}$.
    \label{fig:sigma_K3.0_K0.9_a}
  }
\end{figure}

To end up with the study of diffusion in the thick layer, we have computed the time evolution
of the variance of the histograms for two values of the perturbation parameter.
In Fig.~\ref{fig:sigma_K3.0_K0.9_a} we show  $\sigma^2(L)$, for $0 \leq L\leq 0.2$,
for $\epsilon=10^{-3}$ and $\epsilon=10^{-4}$, respectively with round and square dots.
We have used the variable $L$ in the temporal axe in order to have a time unit comparable
for both $\epsilon$ values. More specifically, $L=0.2$ corresponds to
$n=10^6\times0.2=2\times10^5$ for $\epsilon=10^{-3}$, while
it corresponds to $n=10^8\times0.2=2\times10^7$ for $\epsilon=10^{-4}$.
We observe that there is agreement between both variance evolutions.

\begin{figure}[ht!]
  \includegraphics[width=0.49\textwidth]{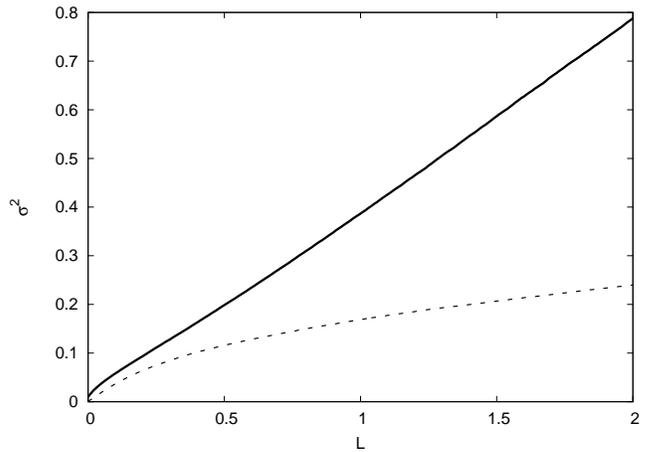}
  \caption{$\sigma^2(L)$, for $0 \leq L\leq 2$ and $\epsilon=10^{-3}$. The upper curve corresponds
    to $K=3.0$ while the lower (dashed) curve corresponds to $K=0.9$.
    \label{fig:sigma_K3.0_K0.9_b}
  }
\end{figure}

The upper curve of Fig.~\ref{fig:sigma_K3.0_K0.9_b} shows the same variance evolution,
only for $\epsilon=10^{-3}$, for a bigger time interval: $0\leq L\leq 2$.
We observe an approximately linear behavior.
This has been corroborated by
making a least square fit of the ansatz $\sigma^2(L)\propto L^w$
and obtaining $w\approx 1.02$ for the time interval $0.5\leq L\leq 2$.
Thus, it can be said that under these circumstances, the global diffusion is
highly close to normal.

\begin{figure}[ht!]
  \includegraphics[width=0.49\textwidth]{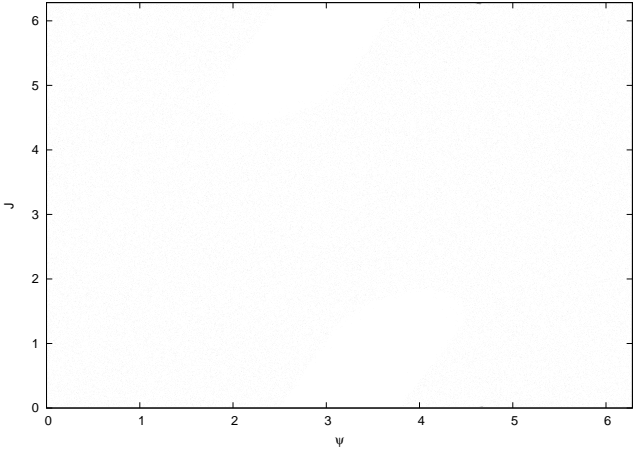}
  \caption{ Snapshot of the phase space $(J,\psi)$ 
    for the map~(\ref{map_Meiss})
    computed after 
    $2\times 10^6$ iterations with $\epsilon=10^{-3}$ and $K=3$; i.e.
    projection onto the $[J,\psi]$--plane of the orbits  
    associated to $\rho_{\epsilon}(I,2\times 10^6)$.
  \label{fig:back_coupling_K3.0}
  }
\end{figure}

In Fig.~\ref{fig:back_coupling_K3.0}, we show a snaphshot of the phase space $(J,\psi)$ computed after
$2\times 10^6$ iterations (largest time used in our
experiments with $\epsilon=10^{-3}$, corresponding to a diffusion time $L=2$), to be compared with Fig.~\ref{fig:SM_K3.0_snapshot_sample_space}.
The similarity of the two pictures suggests that the chaotic layer used to computed the diffusion coefficients is
robust under the effect of $\mathcal{O}(\epsilon)$ perturbations even
after relatively long times.

\section{\label{sec:thin_chaotic_layer} Thin chaotic layer }
\noindent
In this section we will work with a parameter $K=0.9$,
following a similar procedure to that of the previous section.
For this parameter value the area filled by islands of stability
is considerably larger than before.
In the phase space we have different separated chaotic layers, 
so that one has to choose the sample space more carefully 
than in the previous section.

\begin{figure}[ht!]
  \includegraphics[width=0.49\textwidth]{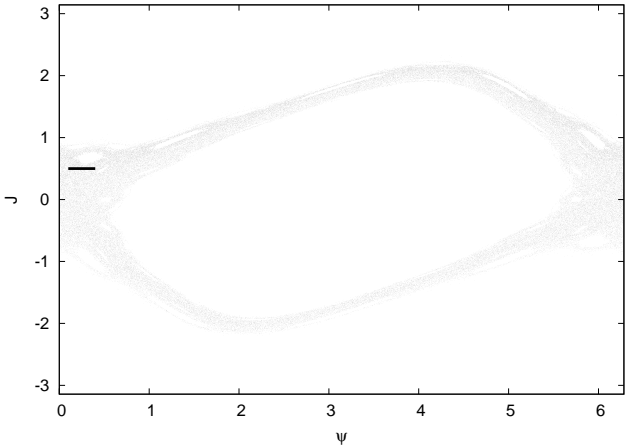}
  \caption{Snapshots of the sample space associated to the SM
    with $K=0.9$: $\mathcal{S}_{0}$ is displayed with a black straight line
    while $\mathcal{S}_{n_s}$ is displayed with dots.
    \label{fig:SM_K0.9_snapshot_sample_space}
    }
\end{figure}

In this opportunity, we choose the $N_r$ seeds of the SM to be
placed along the segment defined by $J=J_0\equiv 0.5$ and $\quad 0.1\leq \psi_0 \leq0.4$.
Thus, we have a new sample space $\mathcal{S}$ defined by the union of the snapshots
\begin{IEEEeqnarray}{ll}
  \label{IC_sample_space_K0.9}
  \mathcal{S}_{n}\equiv \{& (J_{n}^{(k)},\psi_{n}^{(k)})=S^{n}[(J_0, u_k )]:\nonumber \\
  & u_k\in \mathcal{U}(0.1,0.4);\nonumber \\
  & k=1,2,\dots,N_r;\quad N_r=10^5 \},
\end{IEEEeqnarray}
for $n=n_s,n_s+1,\dots,2n_s$ with $n_s=500$.
Therefore, the sample space is the thin chaotic layer
associated to the primary island chain ($J_r=0$).
Fig.~\ref{fig:SM_K0.9_snapshot_sample_space} displays
$\mathcal{S}_{0}$ and $\mathcal{S}_{n_s}$ using a black line and dots,
respectively.

\begin{figure}[ht!]
  \includegraphics[width=0.49\textwidth]{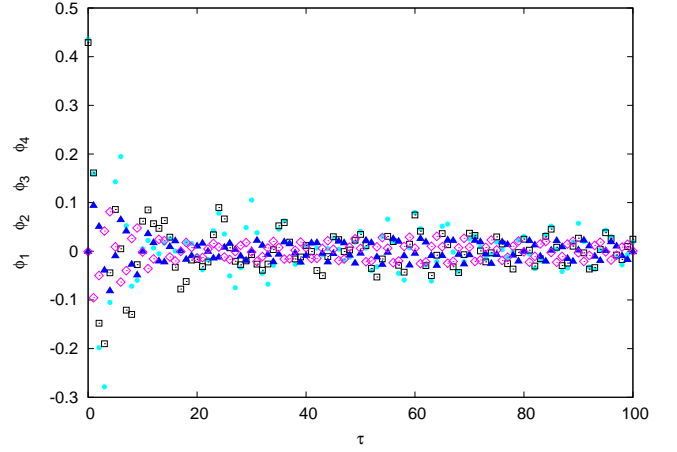}
  \caption{(Color online)
    SM autocorrelation functions: $\phi_i(n_s,n_s+\tau)$ ($n_s=500$) of
    Eq.~(\ref{phi_i_esperanza}), for $i=1,2,3,4$
    using black (square) dots, cyan (round) dots,
    blue (triangle--shaped) dots and magenta (diamond--shaped) dots,
    respectively.
    The sample set is $\mathcal{S}=\bigcup_{n=n_s}^{2n_s}\mathcal{S}_{n}$,
    with $\mathcal{S}_{n}$ defined by Eq.~(\ref{IC_sample_space_K0.9}).
    \label{fig:SM_K0.9_phi_1234}
    }
\end{figure}

In Fig.~\ref{fig:SM_K0.9_phi_1234} we display, for  $0\leq\tau\leq 100$, the values of
$\phi_i(n_s,n_s+\tau)$, for $i=1,2,3,4$
using black (square) dots, cyan (round) dots,
blue (triangle--shaped) dots and magenta (diamond--shaped) dots,
respectively.

Even in this case the autocorrelations fluctuate around zero,
and both $\phi_3$ and $\phi_4$ show the least amplitude of fluctuation.
A distinctive characteristic is the fact that the amplitudes of
$\phi_1(\tau)$ and $\phi_2(\tau)$ are similar to each other.

\begin{figure}[ht!]
  \begin{center}
    \begin{tabular}{c}
      {\includegraphics[width=0.49\textwidth]{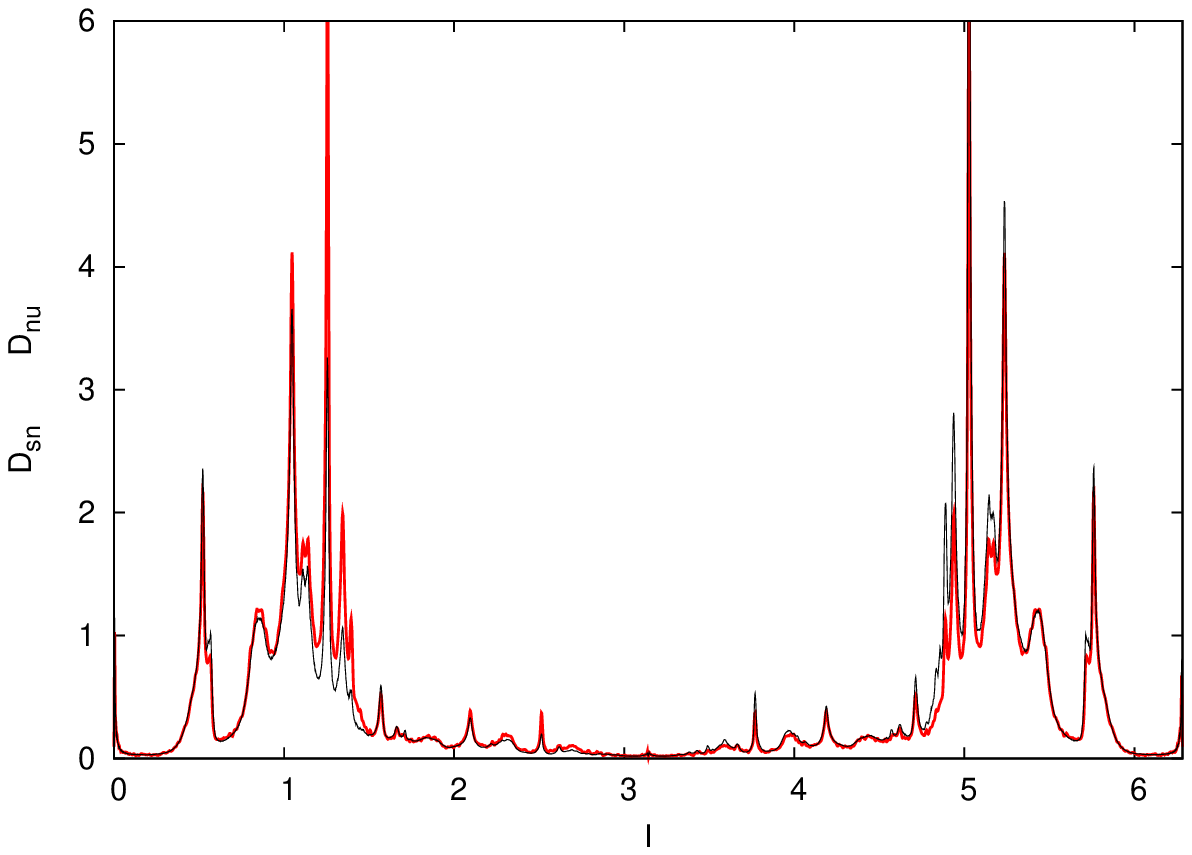}}\\
      { \includegraphics[width=0.49\textwidth]{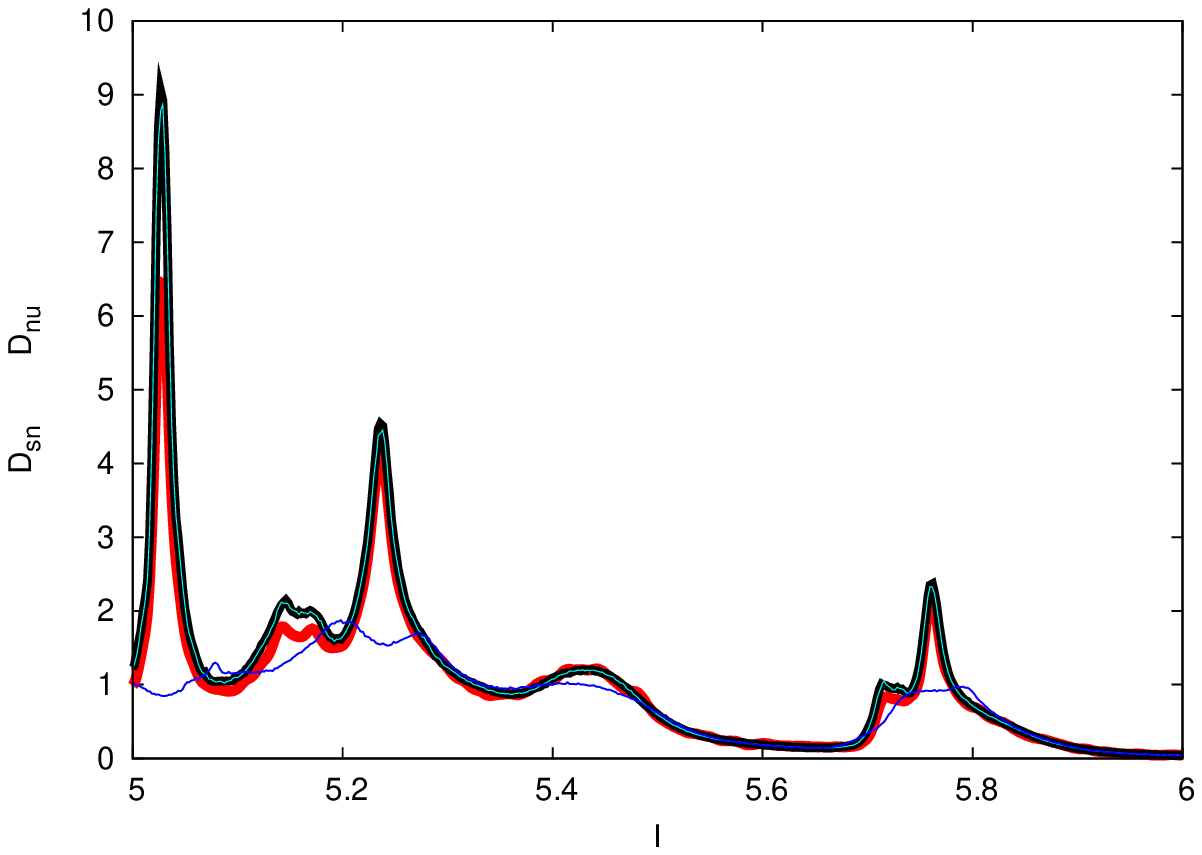}}
    \end{tabular}
    \caption{(Color online) Diffusion coefficients for the symplectic map (\ref{map_Meiss})
    for $K=0.9$ and $\mathcal{S}=\bigcup_{n=n_s}^{2n_s}\mathcal{S}_{n}$
    with $\mathcal{S}_{n}$ defined by Eq.~(\ref{IC_sample_space_K0.9}).
    $\mathcal{D}_{sn}(I;500)$ is shown in red.
    $\mathcal{D}_{nu}(I;500)$ was computed with $\epsilon=10^{-3}, 10^{-5}, 10^{-7}$
    and displayed in colors blue, cyan and black, respectively.
    \label{fig:prediction_coef_K0.9}
  }
  \end{center}
\end{figure}

Fig.~\ref{fig:prediction_coef_K0.9}--top shows, for $3000$ equidistant
values of $I\in [0,2\pi)$, the semi--numerical and the
numerical diffusion coefficients, being
the latter computed for $\epsilon=10^{-7}$.
$\mathcal{D}_{sn}(I;500)$ and $\mathcal{D}_{nu}(I;500)$
are displayed in red and black colors, respectively.
We notice that there is agreement between prediction
and measurement, as in the $K=3$ case.
Fig.~\ref{fig:prediction_coef_K0.9}--bottom shows, for $5\leq I\leq 6$,
the same two curves of the top panel together with
$\mathcal{D}_{nu}(I;500)$ computed with $\epsilon=10^{-3}$ and $\epsilon=10^{-5}$,
in colors blue and cyan, respectively. The result is analogous to
the one obtained in Fig.~\ref{fig:prediction_coef_K3.0}--bottom, regarding
the fact that the method works better for the two smaller
values of $\epsilon$.

This time we have solved numerically the F-P equation with initial actions
distributed according to a $\mathcal{N}(0.75,0.02)$ and the values
of  $\mathcal{D}(\hat{I})$ where those of $\mathcal{D}_{nu}(I;500)$, plotted in
Fig.~\ref{fig:prediction_coef_K0.9}--top.
Besides, we have computed the evolution, for $\epsilon=10^{-3}$, of an ensemble of $N_r$ test particles
with initial conditions given by:

\begin{itemize}
\item $I_0\sim \mathcal{N}(0.75,0.02)$
\item $\theta_0=(1+\sqrt{5})/2\approx 1.618$
\item $(J_0,\psi_0)\in \mathcal{S}_{n_s}$.
\end{itemize}

\begin{figure}[ht!]
  \includegraphics[width=0.49\textwidth]{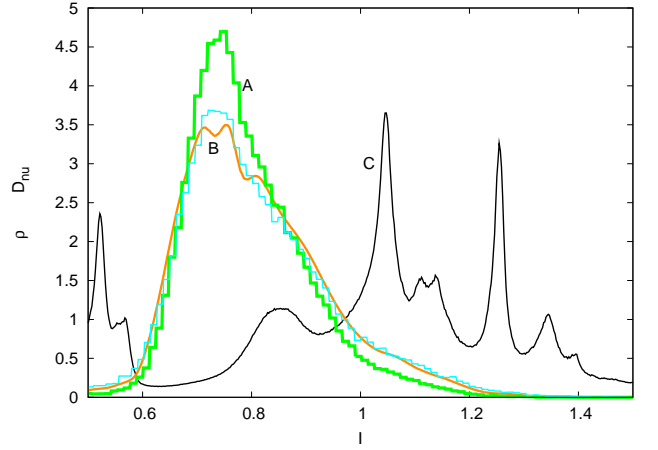}
  \caption{(Color online)
    Distributions for thin layer regime ($K=0.9$):
    $\rho(I,0.03)$, $\rho_{\epsilon}(I,3\times10^4)$ and $\rho_{\epsilon}(I,5\times10^4)$,
    in colors orange (smooth line B),
    green (histogram A)
    and cyan (histogram B), respectively.
    For the histograms the coupling parameter is $\epsilon=10^{-3}$.
    The black line (C) displays the diffusion coefficient used when
    integrating the F-P equation.
  \label{fig:fokker_planck_K0.9_a}
  }
\end{figure}

\begin{figure}[ht!]
  \includegraphics[width=0.49\textwidth]{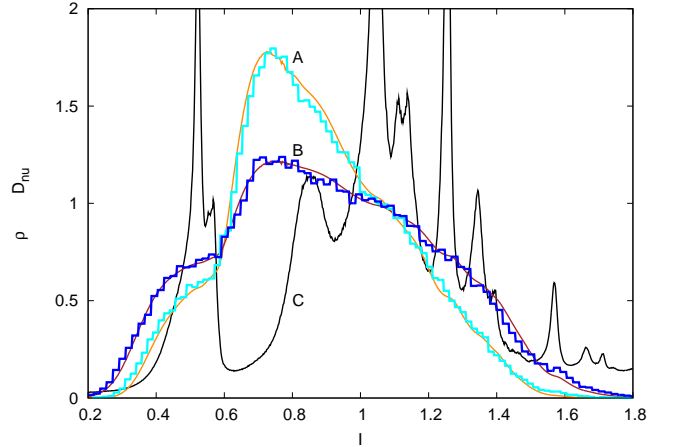}
  \caption{(Color online)
    Distributions:
    $\rho(I,0.10)$, $\rho_{\epsilon}(I,1.9\times 10^5)$,
    $\rho(I,0.19)$ and $\rho_{\epsilon}(I,3.6\times 10^5)$,
    in colors orange (smooth line A), cyan (histogram A),
    brown (smooth line B) and blue (histogram B), respectively.
    For the histograms the coupling parameter is $\epsilon=10^{-3}$.
    The black line (C) displays the diffusion coefficient used when
    integrating the F-P equation.
    \label{fig:fokker_planck_K0.9_b}
    }
\end{figure}

\begin{figure}[ht!]
  \includegraphics[width=0.49\textwidth]{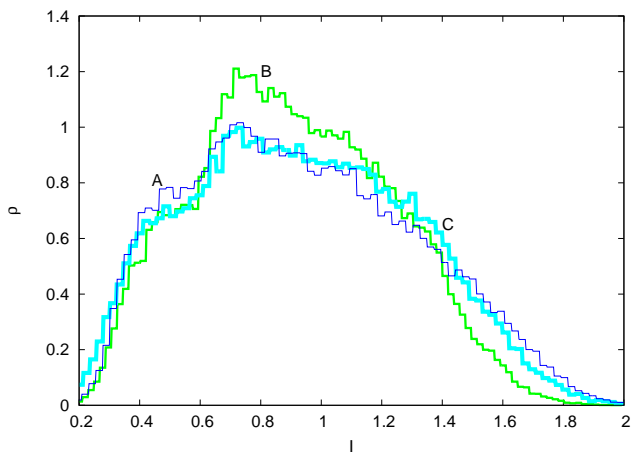}
  \caption{(Color online)
    Histograms for thin layer regime ($K=0.9$):
    $\rho_{\epsilon}(I,4\times10^7)$ ($\epsilon=10^{-4}$),
    $\rho_{\epsilon}(I,4\times10^5)$ ($\epsilon=10^{-3}$) and
    $\rho_{\epsilon}(I,6.5\times10^5)$  ($\epsilon=10^{-3}$),
    in colors blue (A), green (B) and cyan (C), respectively.
  \label{fig:histogram_K0.9_comparacion_pps}
  }
\end{figure}

The orange (smooth line B) and green (histogram A) curves
in Fig.~\ref{fig:fokker_planck_K0.9_a}
represent the functions $\rho(I,0.03)$ and $\rho_{\epsilon}(I,3\times10^4)$,
respectively.
If the relation between the {F-P} and the symplectic times was the one given
by Eq.~(\ref{FP-symplectic_time_relation}), then we would have obtained
that both functions match each other.
Instead, we notice a difference: the {F-P} solution is ahead of,
i.e. more evolved than, the histogram.
In fact, the function that matches
$\rho(I,0.03)$ is $\rho_{\epsilon}(I,5\times10^4)$, which
is shown in cyan color (histogram B),
so that $n=5\times10^4$ corresponds to $L=0.03$ and not to $L=0.05$. Thus, there is
a shift of $\Delta L=0.02$ (for $L=0.03$) in the diffusion time.
Fig.~\ref{fig:fokker_planck_K0.9_b} shows
$\rho(I,0.10)$, $\rho_{\epsilon}(I,1.9\times 10^5)$,
$\rho(I,0.19)$ and $\rho_{\epsilon}(I,3.6\times 10^5)$,
in colors orange (smooth line A), cyan (histogram A),
brown (smooth line B) and blue (histogram B), respectively.
We can deduce that  $\rho(I,0.10)$ matches $\rho_{\epsilon}(I,1.9\times 10^5)$ while
$\rho(I,0.19)$ does the same with $\rho_{\epsilon}(I,3.6\times 10^5)$.
The first pair implies that $\Delta L= 0.19-0.10= 0.09$ for $L=0.10$ and
the second pair implies that $\Delta L= 0.36-0.19 = 0.17$ for $L=0.19$.
As the value of $\Delta L$ is not constant for every time, we have that
the relation between $L$ and $n$ cannot be linear for this $K$ value.
Neither is valid the $\epsilon$--dependence of $L$, as will be shown in
Fig.~\ref{fig:histogram_K0.9_comparacion_pps}.
There it can be seen that the histogram $\rho_{\epsilon}(I,4\times10^7)$ for $\epsilon=10^{-4}$,
shown in blue (A), does not match the concomitant histogram $\rho_{\epsilon}(I,4\times10^5)$
for $\epsilon=10^{-3}$, shown in green (B). Instead, it resembles
the histogram $\rho_{\epsilon}(I,6.5\times10^5)$ for $\epsilon=10^{-3}$, displayed in cyan (C).

We conjecture that this discrepancy is the consequence of the slow correlation 
decaying in the chaotic dynamics with respect to the diffusion time scale 
$\simeq \epsilon^{-2}$ (see also the end of this section). 
Indeed, the numerical simulations point out that the amplitudes of the 
fluctuations of the autocorrelation functions decay much more slowly in the thin, 
than in the thick, layer case. 
This implies that for the smallest K value the system is far away 
from the hypothesis of the (stochastic) averaging
theorem.
We suggest the possibility of a different scaling law between the original time and the diffusion time according to
\begin{equation}
L=\epsilon^2 n^{\alpha(\epsilon,K)}
\label{scalingnew}
\end{equation}
with $\alpha < 1$ and $\lim_{\epsilon\to 0} \alpha(\epsilon, K)=1$
for $K$ values that correspond to sufficiently large chaotic layers.
A simple numerical interpolation from the numerical results with
$\epsilon=10^{-3}$ gives $\alpha\simeq 0.95$, whereas for
$\epsilon=10^{-4}$ we get $\alpha\approx 0.992$ (cfr.
Figs.~\ref{fig:fokker_planck_K0.9_a},~\ref{fig:fokker_planck_K0.9_b}
and~\ref{fig:histogram_K0.9_comparacion_pps}). This anomalous
behavior in the diffusion dynamics is consistent with the local
diffusion that can be observed in Fig.~\ref{fig:sigma_K3.0_K0.9_a},
where a power law with exponent $w\approx0.92$ was fitted for
$\epsilon=10^{-3}$ and $0\leq L \leq 0.1$.

For both layer regimes, the persistence of correlations for every time,
is due to the existence of stability islands because the particles of the chaotic layer,
when getting close enough to such islands, behave regularly, i.e. there the dynamics
is ``locally ordered''.

Finally, let us look at the lower curve
in Fig.~\ref{fig:sigma_K3.0_K0.9_b}, which represents the function
$\sigma^2(L)$, for $0 \leq L\leq 2$ and $\epsilon=10^{-3}$. We infer that the global diffusion character
depends also on the value of the parameter $K$ that determines different chaotic regimes.
In the case of a thin chaotic layer we remark that the variance evolution is not linear so that the
global diffusion is anomalous.
In the interval $0.5\leq L\leq 2$ we have fitted a power law 
with exponent $w\approx  0.53$.
This characterises the macroscopic behavior as a sub-diffusion.
Notwithstanding, the global diffusion behavior is more complex than
a power law and will be considered in a future work.

\begin{figure}[ht!]
  \includegraphics[width=0.49\textwidth]{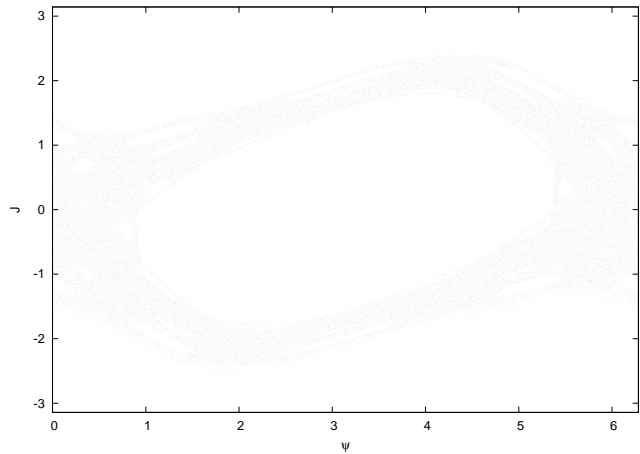}
  \caption{ 
      Snapshot of the phase space $(J,\psi)$ for the 
      map~(\ref{map_Meiss}) computed after
      $2\times 10^6$ iterations with $\epsilon=10^{-3}$ and $K=0.9$; i.e.
      projection onto the $[J,\psi]$--plane of the orbits
      associated to $\rho_{\epsilon}(I,2\times 10^6)$.
    \label{fig:back_coupling_K0.9}
  }
\end{figure}

Moreover in this case the back coupling between the integrable and the chaotic
degree of freedom produces a relevant effect on long time iterations.
In Fig.~\ref{fig:back_coupling_K0.9} we show a
snaphshot of the ensemble iterated with $\epsilon=10^{-3}$, at time $n=2\times 10^6$ ($L=2$) and the
comparison with Fig.~\ref{fig:SM_K0.9_snapshot_sample_space} points out 
a non negligible difference. It will be a subject of future
works to quantify the influence of the back coupling in the
diffusion behaviour, and try to figure out whether the dynamical
origin of the anomalous diffusion is the back coupling, or
is an intrinsic property of the (uncoupled) SM noise, or
it is due to the strong $I$--dependence of the diffusion
coefficient. Nevertheless, the back coupling has absolutely no effect
in the computation of the diffusion coefficient, where
the iteration time used is quite small, i.e. $N=500$.
Plotting a snapshot of the ensemble at this time,
gives a distribution indistinguishable (so not shown)
from the initial conditions from Fig.~\ref{fig:SM_K0.9_snapshot_sample_space}.

\section{\label{sec:conclusion} Conclusion}
\noindent
The main task developed in this article was the study of the diffusion behavior
of a ``{\it quasi--}action'', $I$, of a symplectic 4D {\it a priori}
unstable map.

As a tool to predict the diffusion coefficient we have introduced
a semi--numerical method based on a theorem for stochastically
perturbed Hamiltonian systems. This method allowed us to
estimate the diffusion coefficient for two values of
$K$, which correspond to the situations of
thick and thin chaotic layer, and for a wide
range of the perturbation parameter $\epsilon$ value.
The results are consistent with an $\epsilon^2$--scaling of the
diffusion coefficient which is the same dependence found
in Refs.~\cite{2009CMaPh.290..557G,2010CeMDA.107..115L},
also for an {\it a priori} unstable system.

We worked with chaotic test particles, because the ensemble of orbits
$\mathcal{S}$ was chosen in such a way that the
initial conditions in the  $[J,\psi]$--plane belonged to a chaotic layer.

Even though for the symplectic map we did not use an analytic expression,
the semi--numerical diffusion coefficient is a prediction because
it uses as input data only the unperturbed  motion of the integrable part
(the free rotator) and the numerical autocorrelation function of the
perturbation (SM).

Regarding the behavior of the action distribution, $\rho_\epsilon(I,n)$,
we have found two different numerical results, depending on the chaotic regime.

On the one hand, in the thick layer case, the distribution follows a F-P equation.
Besides, the time scaling relation, $L=\epsilon^2 n$, proposed in the stochastic theorem,
between the (slow) F-P time $L$, and the symplectic time $n$, turns out to be correct.

On the other hand, in the thin layer case, the distribution presents
a ``time delay'' respect to the limit F-P solution. This discrepancy
is sensible because, in this regime, the SM ``noise'' is too far
away from the conditions asked for in the stochastic theorem, mainly
regarding the necessity that the autocorrelation function must decay
to zero. We conjecture that if the correlation function is not
decaying sufficiently fast with respect to the diffusion time scale
$\epsilon^{-2}$, we have an anomalous scaling for fixed $K$ values,
between the initial time and the diffusion time according to
$L=\epsilon^2 n^{\alpha(\epsilon,K)}$ where $\alpha< 1$ and
$\lim_{\epsilon\to 0}\alpha(\epsilon,K)=1$.
This conjecture will be analyzed in a future work together
with the character of global diffusion.
In the thick layer regime, the diffusion behaved normally, while in the other case,
it behaved sub--normally.

The applicability of the presented stochastic approach goes beyond the
particular 2.5DoF Hamiltonian system chosen in this work. Specially, it
would allow to characterize and quantify diffusion in Hamiltonian
systems that model perturbed simple nonlinear resonances.

We have also studied diffusion of a system built by coupling the free rotator with
an ensemble of damped stochastic harmonic oscillators, finding
a diffusion coefficient which differs considerably respect to its deterministic counterpart.
Moreover we have justified why it is not possible to
model a SM--driven diffusion with the one driven by a stochastic rotator.

In resume, we have given empirical evidence of the fact that a chaotic layer can
act as a stochastic pump when coupled to some integrable system. After this results,
we consider that the other way in which the chaotic layer is usually named after, i.e.
{\it stochastic layer}, is phenomenologically suitable.

\begin{acknowledgments}
We acknowledge the anonymous referees for their suggestions and constructive criticism, which helped to correct and improve this work.
The stay of Mestre at the Physics Department of the University of Bologna
was fully supported by a grant from the Erasmus Mundus External
Cooperation Window Lot 16 Programme, EADIC, financed by the European Commission.
Besides, Cincotta, Giordano and Mestre
were supported with grants from the
Consejo Nacional de Investigaciones Cient\'ificas y
T\'ecnicas de la Rep\'ublica Argentina (CONICET).
This research has made use of NASA's Astrophysics Data System Bibliographic Services.
\end{acknowledgments}

\appendix
\section{\label{app:autocorrel_damped_harm_osc}Stationary autocorrelation function of the damped harmonic
stochastic oscillator}
We will need two properties of the (Ito) stochastic integral of deterministic functions,
that can be demonstrated from the basic definition found in elementary textbooks
of the subject (for example see Refs.~\cite{2004Gardiner,1973ArnoldLudwig}).
These are the {\it mean value} formula:
\begin{equation*}
  E\left[\int_{t_0}^{t}f(z) dW_z\right]=0
\end{equation*}
and the {\it autocorrelation} formula:
\begin{IEEEeqnarray*}{l}
  E\left[\int_{t_0}^{t}f(z) dW_z \int_{t_0}^{t'}g(z) dW_z\right]\\
  =~\int_{t_0}^{min(t,t')} f(z)g(z) dz,
\end{IEEEeqnarray*}
valid for arbitrary continuous functions $f(z)$ and $g(z)$ and arbitrary
times $t$ and $t'$.

From Eq.~(\ref{solution_damped_harm_osc}),  the soluction $\xi_t$ can be written
as the sum of a deterministic term, denoted as $\xi_t^{(d)}$, and a
stochastic one, denoted as $\xi_t^{(s)}$, according to the following definitions:
\begin{IEEEeqnarray*}{ll}
  \xi_t^{(d)} =&~\frac{\lambda \xi_0+ 2 v_0}{2\omega_1}
  \mathrm{e}^{-\frac{\lambda}{2}t} \sin(\omega_1t) +
  \xi_0  \mathrm{e}^{-\frac{\lambda}{2}t} \cos(\omega_1t) \\
  \xi_t^{(s)} = &~\frac{\sqrt{c}}{\omega_1}
  \int_0^t  \mathrm{e}^{-\frac{\lambda}{2}(t-z)}\sin(\omega_1(t-z))dW_z.
\end{IEEEeqnarray*}
The mean value formula for $f(z)\equiv h(z;t)=\mathrm{e}^{-\frac{\lambda}{2}(t-z)}\sin(\omega_1(t-z))$
($t$ works as a parameter respect to the integral) implies that:
$E[\xi_t^{(s)}]=0$ so that
\begin{equation}
  \label{app_A_autocorrelation_exact}
  E[\xi_t \xi_{t'}]= \xi_t^{(d)} \xi_{t'}^{(d)} + E[\xi_t^{(s)} \xi_{t'}^{(s)}].
\end{equation}
Considering also $g(z)\equiv h(z,t')$ and using the autocorrelation formula we have that:
\begin{IEEEeqnarray*}{ll}
E[\xi_t^{(s)} \xi_{t'}^{(s)}]=\frac{c}{\omega_1^2}\int_0^{min(t,t')}h(z;t)h(z;t')dz=\\
\frac{c}{\omega_1^2}\int_0^{min(t,t')}
\mathrm{e}^{-\frac{\lambda}{2}(t+t'-2z)}\sin\big(\omega_1(t-z)\big)\sin\big(\omega_1(t'-z)\big) dz.
\end{IEEEeqnarray*}
Integrating this expression and replacing $t'=t+\tau$, for $\tau\geq 0$, we
obtain that the stochastic contibution to the autocorrelation function is given by:
\begin{IEEEeqnarray*}{ll}
  E[\xi_t^{(s)} \xi_{t+\tau}^{(s)}]=&\frac{c}{4\omega_1^2}\mathrm{e}^{-\frac{\lambda}{2}\tau}
  \left\{
  \left(
  \frac{\mathrm{e}^{\mathrm{i}\omega_1\tau}+\mathrm{e}^{-\mathrm{i}\omega_1\tau} }{\lambda}
  \right)
  (1-\mathrm{e}^{-\lambda t} )\right.\\
  &- \frac{ \mathrm{e}^{\mathrm{i}\omega_1\tau} } { (\lambda -\mathrm{i}2\omega_1) }
  (1-\mathrm{e}^{-\lambda t} \mathrm{e}^{\mathrm{i}2\omega_1t})\\
  & \left. - \frac{ \mathrm{e}^{-\mathrm{i}\omega_1\tau} } { (\lambda +\mathrm{i}2\omega_1) }
  (1-\mathrm{e}^{-\lambda t} e^{-\mathrm{i}2\omega_1t})
 \right\}.
\end{IEEEeqnarray*}
Then, the following asymptotic behavior is satisfied:
\begin{IEEEeqnarray}{l}
  \label{app_A_asymptotic_autocorrel_stochastic}
  \lim_{t\rightarrow +\infty} E[\xi_t^{(s)} \xi_{t+\tau}^{(s)}]\nonumber \\
  =\frac{c}{2 \lambda \omega^2}\mathrm{e}^{-\frac{\lambda}{2}\tau}
 \{\cos(\omega_1\tau)+\frac{\lambda}{2\omega_1}\sin(\omega_1\tau)\}.
\end{IEEEeqnarray}
On the other side, it can be proven that the deterministic part does not
contribute asymptotically because
\begin{equation}
  \label{app_A_asymptotic_autocorrel_deterministic}
  \lim_{t\rightarrow +\infty} \xi_t^{(d)} \xi_{t+\tau}^{(d)} = 0
\end{equation}

From Eqs.~(\ref{app_A_autocorrelation_exact}),~(\ref{app_A_asymptotic_autocorrel_stochastic})
and~(\ref{app_A_asymptotic_autocorrel_deterministic}) we obtain the desired expression:
\begin{IEEEeqnarray*}{ll}
  \phi_s(\tau)\equiv &~\lim_{t\rightarrow +\infty} E[\xi_t \xi_{t+\tau}]\\
  &=~\frac{c}{2 \lambda \omega^2}\mathrm{e}^{-\frac{\lambda}{2}\tau}
  \{\cos(\omega_1\tau)+\frac{\lambda}{2\omega_1}\sin(\omega_1\tau)\}.
\end{IEEEeqnarray*}

\section{\label{app:spect_dens_damped_harm_osc}Spectral density of the damped harmonic stochastic oscillator}
According to the definition of the spectral density given in Eq.~(\ref{spectral_dens})
and rewritting the trigonometric terms in the autocorrelation function
with exponential functions, we obtain:
\begin{IEEEeqnarray}{l}
  \label{app_B_spectral_density_Gpm}
   \tilde{\phi}_s(\nu) =   \frac{c}{4\lambda \omega^2} \times \nonumber \\
   \left\{
   \left(1-\mathrm{i}\frac{\lambda}{2\omega_1}\right) G_{(+)}
   +
   \left(1+\mathrm{i}\frac{\lambda}{2\omega_1}\right) G_{(-)}
   \right\}
\end{IEEEeqnarray}
where we have introduced the series:
\begin{IEEEeqnarray*}{ll}
  G_{(\pm)}=&~\sum_{m=-\infty}^{+\infty}
  \exp
  \left[-\frac{\lambda}{2}|m|+\mathrm{i}(\nu m \pm \omega_1|m|)\right]\\
  =&~1+\sum_{m=1}^{+\infty} {[p_{(\pm)}]}^m +\sum_{m=1}^{+\infty} {[q_{(\pm)}]}^m,
\end{IEEEeqnarray*}
with
\begin{IEEEeqnarray}{ll}
  \label{app_B_def_p_and_q}
  p_{(\pm)}=&~\exp\left[-\frac{\lambda}{2}+\mathrm{i}(\pm\omega_1-\nu)\right],\nonumber \\
  q_{(\pm)}=&~\exp\left[-\frac{\lambda}{2}+\mathrm{i}(\pm\omega_1+\nu)\right].
\end{IEEEeqnarray}

As it happens that $|p_{(\pm)}|=|q_{(\pm)}|=\mathrm{e}^{-\frac{\lambda}{2}} < 1$ ($\forall \lambda>0$),
the series converge to:
\begin{equation}
  \label{app_B_G_pq}
  G_{(\pm)}=1+ \frac{p_{(\pm)}}{1-p_{(\pm)}}+ \frac{q_{(\pm)}}{1-q_{(\pm)}}.
\end{equation}
Introducing Eqs.~(\ref{app_B_def_p_and_q}) and~(\ref{app_B_G_pq})
into Eq.~(\ref{app_B_spectral_density_Gpm}) and applying some arithmetical and
trigonometrical properties we arrive at a real--valued expression for the spectral
density of the harmonic noise:
\begin{equation}
  \label{spect_dens_damped_harm_osc_1}
  \tilde{\phi_s}(\nu)=\frac{c}{4\lambda\omega^2}
  \left[F_1(\nu;\lambda,\omega_1)+\frac{\lambda}{2\omega_1}F_2(\nu;\lambda,\omega_1)\right],
\end{equation}
where the auxiliary functions $F_k$ ($k=1,2$) are given by:
\begin{IEEEeqnarray}{ll}
  \label{spect_dens_damped_harm_osc_2}
  F_1 \equiv &~ \frac{Z~X_{(+)}+U_{(+)}Y_{(+)}}{|Q_{(+)}|^2}
  + \frac{Z~X_{(-)}+U_{(-)}Y_{(-)}}{|Q_{(-)}|^2} \nonumber \\
  F_2 \equiv &~ \frac{Z~Y_{(+)}+U_{(+)}X_{(+)}}{|Q_{(+)}|^2}
  - \frac{Z~Y_{(-)}+U_{(-)}X_{(-)}}{|Q_{(-)}|^2},
\end{IEEEeqnarray}
with:
\begin{widetext}
\begin{IEEEeqnarray}{lll}
  \label{spect_dens_damped_harm_osc_3}
  X_{(\pm)}&\equiv&~1 +\mathrm{e}^{-\lambda}\cos(2\omega_1)
  -\mathrm{e}^{-\frac{\lambda}{2}}[\cos(\pm\omega_1-\nu)+\cos(\pm\omega_1+\nu)],\nonumber \\
  Y_{(\pm)}&\equiv&~\pm \mathrm{e}^{-\lambda}\sin(2\omega_1)
  -\mathrm{e}^{-\frac{\lambda}{2}}[\sin(\pm\omega_1-\nu)+\sin(\pm\omega_1+\nu)],\nonumber \\
  Z~&\equiv&~1-\mathrm{e}^{-\lambda}\cos(2\omega_1),\nonumber \\
  U_{(\pm)}&\equiv& \mp \mathrm{e}^{-\lambda}\sin(2\omega_1),\nonumber \\
  Q_{(\pm)}&\equiv& X_{(\pm)}+\mathrm{i} Y_{(\pm)}.
\end{IEEEeqnarray}
\end{widetext}


%

\end{document}